\pdfoutput=1

\documentclass[11pt,twoside,a4paper,cmspaper,final,collab]{cms-tdr}
\def\svnVersion{475474P}\def\cmsCernNoTag{CERN-EP-2018-079}\def\cmsCernDate{\today}\def\cmsMessage{Published in the Journal of High Energy Physics as \href{http://dx.doi.org/10.1007/JHEP09(2018)065}{\doi{10.1007/JHEP09(2018)065.}}}

\begin{document}\cmsNoteHeader{SUS-17-005}

\hyphenation{had-ron-i-za-tion}
\hyphenation{cal-or-i-me-ter}
\hyphenation{de-vices}
\RCS$HeadURL: svn+ssh://svn.cern.ch/reps/tdr2/papers/SUS-17-005/trunk/SUS-17-005.tex $
\RCS$Id: SUS-17-005.tex 474127 2018-09-06 14:55:40Z bargassa $

\newlength\cmsFigWidth
\ifthenelse{\boolean{cms@external}}{\setlength\cmsFigWidth{0.85\columnwidth}}{\setlength\cmsFigWidth{0.4\textwidth}}
\ifthenelse{\boolean{cms@external}}{\providecommand{\cmsLeft}{top\xspace}}{\providecommand{\cmsLeft}{left\xspace}}
\ifthenelse{\boolean{cms@external}}{\providecommand{\cmsRight}{bottom\xspace}}{\providecommand{\cmsRight}{right\xspace}}
\providecommand{\cmsTable}[1]{\resizebox{\textwidth}{!}{#1}}
\providecommand{\NA}{\ensuremath{\text{---}}}

\newcommand{\Iabs}{\ensuremath{I_\text{abs}}}
\newcommand{\Irel}{\ensuremath{I_\text{rel}}}
\newcommand{\CT}{\ensuremath{C_\mathrm{T}}\xspace}
\newcommand{\Zg}{\ensuremath{\cPZ/\gamma^*}\xspace}
\newcommand{\dxy}{\ensuremath{d_{xy}}\xspace}
\newcommand{\dz}{\ensuremath{d_{z}}\xspace}
\newcommand{\mt}{\ensuremath{M_{\mathrm{T}}}}
\newcommand{\ctone}{\ensuremath{C_{\mathrm{T1}}}}
\newcommand{\cttwo}{\ensuremath{C_{\mathrm{T2}}}}
\newcommand{\ct}{\ensuremath{C_{\mathrm{T}}}}
\newcommand{\njets}{\ensuremath{N_{\text{jets}}}}
\newcommand{\nbl}{\ensuremath{N(\cPqb^{\text{loose}})}}
\newcommand{\ptl}{\ensuremath{\pt(\ell)}\xspace}
\newcommand{\etl}{\ensuremath{\eta(\ell)}\xspace}
\newcommand{\chgl}{\ensuremath{Q(\ell)}}
\newcommand{\ptisr}{\ensuremath{\pt(\mathrm{ISR})}}
\newcommand{\ptb}{\ensuremath{\pt(\cPqb)}}
\newcommand{\drLB}{\ensuremath{\Delta R (\ell, \cPqb)}}
\newcommand{\bdisc}{\ensuremath{D(\cPqb)}}
\newcommand{\Nb}{\ensuremath{N_\cPqb}\xspace}
\newcommand{\wjets}{\ensuremath{\PW}+jets}
\newcommand{\zinv}{\ensuremath{\cPZ(\to\nu\nu)+\text{jets}}\xspace}
\newcommand{\mmplane}{\ensuremath{ (  {m}(\stp),   {m}(\lsp) )}\xspace}
\newcommand{\lsp}{\PSGczDo}
\newcommand{\chg}{\PSGcpmDo}
\newcommand{\stp}{\ensuremath{\PSQt_{1}}\xspace}
\newcommand{\stpb}{\ensuremath{{\PASQt}}_{1}\xspace}
\newcommand{\dm}{\ensuremath{\Delta {m}}}

\cmsNoteHeader{SUS-17-005}

\title{Search for top squarks decaying via four-body or chargino-mediated modes in single-lepton final states in proton-proton collisions at $\sqrt{s} = 13\TeV$}

\date{\today}

\abstract{
A search for the pair production of the lightest supersymmetric partner of
the top quark (\stp) is presented. The search focuses on a compressed
scenario where the mass difference between the top squark and the lightest
supersymmetric particle, often considered to be the lightest neutralino
(\lsp), is smaller than the mass of the \PW\ boson. The proton-proton
collision data were recorded by the CMS experiment at a centre-of-mass
energy of 13\TeV, and correspond to an integrated luminosity of
35.9\fbinv. In this search, two decay modes of the top squark are
considered: a four-body decay into a bottom quark, two additional
fermions, and a \lsp; and a decay via an intermediate chargino. Events are
selected using the presence of a high-momentum jet, significant missing
transverse momentum, and a low transverse momentum electron or muon. Two
analysis techniques are used, targeting different decay modes of the \stp:
a sequential selection and a multivariate technique. No evidence for the
production of top squarks is found, and mass limits at 95\% confidence
level are set that reach up to 560\GeV, depending on the $m(\stp) - m(\lsp)$
mass difference and the decay mode.
}

\hypersetup{
pdfauthor={CMS Collaboration},
pdftitle={Search for top squarks decaying via four-body or chargino-mediated modes in single-lepton final states in proton-proton collisions at sqrt(s) = 13 TeV},
pdfsubject={CMS},
pdfkeywords={CMS, physics, SUSY}}

\maketitle

\section{Introduction}
\label{s:intro}

Searches for new phenomena, in particular supersymmetry
(SUSY)~\cite{Martin,SUSY0,SUSY1,SUSY2,SUSY3,SUSY4}, are among the main
objectives of the physics programme at the CERN LHC. Supersymmetry, which
is one of the most promising extensions of the standard model (SM),
predicts superpartners of SM particles, where the spin of each new
particle differs by one-half unit with respect to its SM counterpart. If
$R$-parity~\cite{Farrar:1978xj}, a new quantum number, is conserved,
supersymmetric particles would be pair-produced and their decay chains
would end with the lightest supersymmetric particle (LSP). Supersymmetric
models can offer solutions to several shortcomings of the SM, in
particular those related to the explanation of the mass hierarchy of
elementary particles~\cite{hierarchy1,hierarchy2} and to the presence of
dark matter in the universe. The search for SUSY has special interest in
view of the recent discovery of the Higgs boson~\cite{HgDisc2,HgDisc1,HgDisc3} as it naturally solves the problem of
quadratically divergent loop corrections to the mass of the Higgs boson by
associating with each SM particle a supersymmetric partner having the same
gauge quantum numbers. In many models of SUSY, the lightest neutralino
\PSGczDo is the LSP and, being neutral and weakly interacting, would match
the characteristics required for a dark matter particle.

\begin{figure}[!htbp]
\centering
\includegraphics{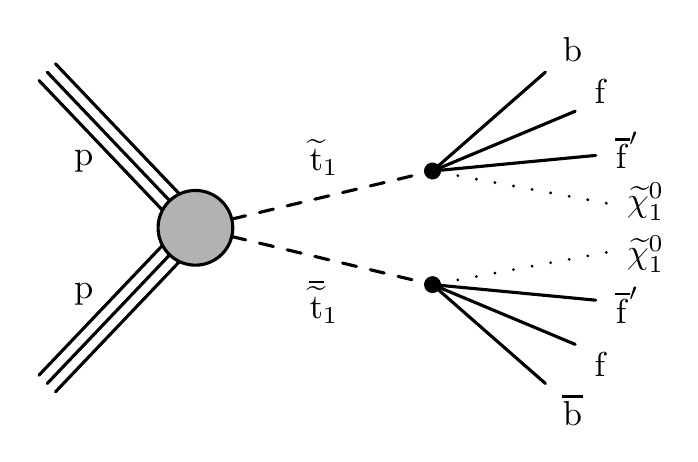}
\includegraphics{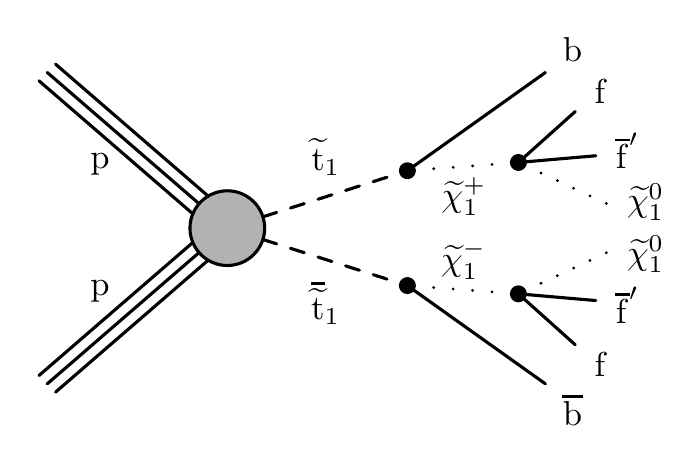}
\caption{Top squark pair production at the LHC with four-body (left)
        or chargino-mediated (right) decays.}
\label{fig:models}
\end{figure}

Supersymmetry predicts a scalar partner for each SM left- and right-handed
fermion. When SUSY is broken, the scalar partners acquire masses different
from those of their SM counterparts, and the mass splitting between the
two squark mass eigenstates is proportional to the mass of their SM
partner. Given the large mass of the top quark, this splitting can be the
largest among all squarks. Therefore the lightest supersymmetric partner
of the top quark, the \stp, is often the lightest squark. Furthermore, if
SUSY is a symmetry of nature, cosmological observations may suggest the
lightest top squark to be almost degenerate with the LSP~\cite{Coannihilation}.
This motivates the search for a four-body \stp
decay: $\stp \to \cPqb \mathrm{f} \overline{\mathrm{f}}^{\,\prime} \PSGczDo$, where
the fermions f and $\overline{\mathrm{f}}^{\,\prime}$ can be either quarks or
leptons. Here, due to the small mass difference between the \stp and the
\PSGczDo, two-body ($\stp \to \cPqt \PSGczDo$, $\stp \to \cPqb \tilde{\chi}_{1}^{+}$)
and three-body ($\stp \to \cPqb \PWp \PSGczDo$) decays of the lightest top squark are
kinematically forbidden, and the two-body ($\stp\to\cPqc\PSGczDo$)
decay can be suppressed depending on the details of the model.
Alternatively, the decay $\stp \to \cPqb \tilde{\chi}_{1}^{+} \to \cPqb
\mathrm{f} \overline{\mathrm{f}}^{\,\prime} \PSGczDo$ is possible if the mass of the
lightest chargino is lower than the top squark mass.
Figure~\ref{fig:models} represents the production of a pair of \stp
followed by a four-body or chargino-mediated decay in simplified models~\cite{sms}.

In this paper, we describe a search for pair production of the \stp in
proton-proton (\Pp\Pp) collisions at the LHC at $\sqrt{s}=13\TeV$, where
each top squark can decay either directly, or via a chargino, into the
$\cPqb \mathrm{f} \overline{\mathrm{f}}^{\,\prime} \PSGczDo$ final state. A 100\%
branching fraction for each decay is assumed when interpreting the results~\cite{sms}.
The final states considered contain jets, missing transverse
momentum (\ptmiss), and exactly one lepton, which can be either an electron
or a muon, originating from the decay of the top squark or the chargino,
depending on the considered decay scenario. The lepton can be efficiently
reconstructed and identified with transverse momentum (\pt) as low as 5.0
and 3.5\GeV for electrons and muons, respectively. In this search, we
expand the result of a previous CMS search in \Pp\Pp\ collisions at
$\sqrt{s} = 8 \TeV$~\cite{Khachatryan:2015pot} by including the
single-electron final state and lowering the \pt thresholds for leptons.
Moreover, two different approaches are used in this analysis. A signal
selection based on sequentially applied requirements on several
discriminating variables (CC) has been designed to provide good
sensitivity over a wide range of kinematic signatures corresponding to
different \mmplane mass hypotheses and different \stp decay modes. The CC
approach is applied to the four-body and chargino-mediated \stp decay
scenarios. In addition, a multivariate analysis (MVA) followed by a
counting experiment approach is used for the signal selection. Applied to
the four-body scenario, this approach exploits the correlations between
discriminating variables and is adapted for different $\dm = m(\stp) - m(\lsp)$
kinematic regions, thus optimizing the search across the
\mmplane space and improving upon the sensitivity of the CC approach for
this scenario. Both approaches are based on a nearly identical
preselection.

Other results in the single-lepton final state and for both the four-body
and chargino-mediated \stp decays were reported by ATLAS at $\sqrt{s} = 8
\TeV$~\cite{Aad:2014kra} and $\sqrt{s} = 13 \TeV$~\cite{Aaboud:2017aeu}.
Other final states at $\sqrt{s} = 13 \TeV$ were investigated by ATLAS and
CMS for all-hadronic events~\cite{Aaboud:2017phn,Sirunyan:2017wif} and for
final states with two isolated
leptons~\cite{Aaboud:2017nfd,Sirunyan:2018iwl}, respectively.

\section{Detector and object definition}
\label{s:objects}

The central feature of the CMS apparatus is a superconducting solenoid of
6\unit{m} internal diameter, providing a magnetic field of 3.8\unit{T}.
Within the solenoid volume are a silicon pixel and strip tracker, a lead
tungstate crystal electromagnetic calorimeter (ECAL), and a brass and
scintillator hadron calorimeter (HCAL), each composed of a barrel and two
endcap sections. Forward calorimeters extend the pseudorapidity coverage
provided by the barrel and endcap detectors. The silicon tracker measures
charged particles within the pseudorapidity range $\abs{\eta}<2.5$. Muons
are detected in gas-ionization chambers embedded in the steel flux-return
yoke outside the solenoid. The detector is nearly hermetic, allowing for
momentum balance measurements in the plane transverse to the beam axis.
Events are selected for further analysis by a two-tier trigger system that
uses custom hardware processors to make a fast initial selection, followed
by a more detailed selection executed on a dedicated processor farm. A
more detailed description of the CMS detector can be found in
Ref.~\cite{ref:CMS}.

This analysis utilizes the CMS particle-flow (PF)
algorithm~\cite{CMS-PRF-14-001} to reconstruct and identify PF candidates
such as leptons (electrons and muons), photons, and charged and neutral
hadrons. The reconstructed vertex with the largest value of summed
physics-object $\pt^2$ is taken to be the primary \Pp\Pp\ interaction vertex
(PV). The physics objects are the jets, clustered using a jet finding
algorithm~\cite{Cacciari:2008gp,Cacciari:2011ma} with the tracks assigned
to the vertex as inputs, and the associated missing transverse momentum,
taken as the negative vector sum of the \pt of those jets.

The electron candidates are reconstructed from energy depositions in the
ECAL and from tracks in the inner tracker obtained using the Gaussian-sum
filter~\cite{Khachatryan:2015hwa}. The misidentification of electrons is
reduced by requiring additional constraints on the shape of the
electromagnetic shower in the ECAL, the quality of the match between the
trajectory of the track and the ECAL energy deposit, and the relative HCAL
deposition in the electron direction. For reconstructing muons the tracks
in both the silicon tracker and the muon system are
used~\cite{Chatrchyan:2012xi}. The number of measurements in the tracker
and muon system and the quality of the track fit are used to reduce the
misidentification rate of muons. In order to select leptons ($\ell=\Pe$ or
$\mu$) from the primary interaction, the point of closest approach to the
PV of tracks associated with the lepton is required to have transverse
component $\abs{\dxy}<0.02\unit{cm}$ and longitudinal component
$\abs{\dz}<0.1\unit{cm}$ with respect to the PV. In order to suppress the
selection of nonprompt leptons, which may arise from jets produced in
association with the invisible decay of a \Z boson, multijet production,
or \wjets\ and \ttbar events with a lost lepton, selected leptons are
required to be isolated from jet activity by using a combination of
absolute and relative isolation variables. The absolute isolation (\Iabs)
of the lepton is defined as the scalar sum of the \pt of PF candidates
within a cone size of $R \equiv \sqrt{\smash[b]{(\Delta\phi)^2+(\Delta\eta)^2}} =
0.3$. The leptons and charged PF candidates not associated with the PV are
not included in the sum. The contribution of the neutral particles from
simultaneous \Pp\Pp\ collisions (pileup) is estimated according to the method
described in Ref.~\cite{Khachatryan:2015hwa}, and subtracted from \Iabs.
The relative isolation (\Irel) of a lepton is defined as the ratio of
lepton \Iabs~to the lepton \pt. The electrons and muons are then required
to satisfy $\Iabs<5\GeV$ for $\pt(\ell)<25\GeV$ and $\Irel<0.2$ for
$\pt(\ell)>25\GeV$. This combined isolation criterion allows for a more
uniform selection efficiency of leptons as a function of lepton \pt.
Finally, the selected electrons and muons are also required to have \pt
above $5.0\GeV$ and $3.5\GeV$ and $\abs{\eta}<2.5$ and $2.4$, respectively.
Tau leptons with a hadronic decay are reconstructed from the PF candidates
using the ``hadrons-plus-strips'' algorithm~\cite{Khachatryan:2015dfa},
which achieves an efficiency of 50--60\%. The tau candidates are required
to satisfy $\abs{\eta}<2.4$.

The jets used in this analysis are reconstructed by clustering PF
candidates using the anti-\kt algorithm~\cite{Cacciari:2008gp} with a
distance parameter of 0.4. The missing transverse momentum vector,
\ptvecmiss, in the event is defined as the negative of the vectorial sum
of the transverse momenta of all the PF candidates in the event with its
magnitude denoted as \ptmiss. The pileup contribution to the jet momenta is
partially taken into account by subtracting the energy of charged hadrons
originating from a vertex other than the PV. The jet momenta are further
calibrated to account for contributions from neutral pileup and any
inhomogeneities of detector response~\cite{Khachatryan:2016kdb}. The jets
have a threshold of $\pt > 30\GeV$ and are required to have $\abs{\eta}<2.4$.

Jets originating from bottom (\cPqb) quarks are identified (``tagged'') as
``\cPqb\ jets'' using the combined secondary vertex
algorithm~\cite{Chatrchyan:2012jua,Sirunyan:2017ezt}, which takes
advantage of MVA techniques. The medium working point of this algorithm is
used in the CC search, which has a probability of about 1\% to misidentify
a light quark jet as a \cPqb\ jet while correctly identifying a \cPqb\ jet
with an efficiency of about 65\%. The same figures for the loose working
point, which is used in the MVA search, are 10\% and 80\%, respectively.

\section{Samples and preselection}

\subsection{Data and simulated samples}
\label{s:samples}

The searches described in this paper are performed using data from \Pp\Pp\
collisions recorded in 2016 by the CMS experiment at the LHC at a
centre-of-mass energy of 13\TeV corresponding to an integrated luminosity
of 35.9\fbinv. Events in the search are collected based on \ptmiss
triggers with thresholds ranging between 90 and 120\GeV. Additional
control samples used for estimating backgrounds are selected by
single-lepton triggers with \pt thresholds of 24 and 27\GeV for muons and
electrons, respectively.

In this analysis, Monte Carlo (MC) simulation samples of SM processes are
used to relate background yields in control and signal regions, to
validate the background estimation methods based on data, and to predict
contributions from rare processes. Simulated samples are produced using
multiple generators. The main background samples, namely \wjets, \ttbar,
and \Zg are generated at leading order (LO) by
\MGvATNLO~2.3.3~\cite{Alwall:2014hca}. Next-to-leading order (NLO)
simulations with the \POWHEG~v2.0~\cite{Alioli:2009je} and
\POWHEG~v1.0~\cite{Re:2010bp} generators are used for single top quark
production and the associated \cPqt\PW\ production, respectively. Diboson
events are simulated at NLO with \MGvATNLO~2.3.3 and \POWHEG~v2.0. The LO
and NLO NNPDF3.0~\cite{Ball:2014uwa} parton distribution functions (PDFs)
are used consistently with the order of the matrix element calculation in
the generated events. Hadronization and showering of events in all
generated samples have been simulated using
\PYTHIA~8.212~\cite{Sjostrand:2006za,Sjostrand:2007gs} with the
CUETP8M1~\cite{CMS-PAS-GEN-14-001} tune for the underlying event. The
response of the CMS detector is modelled using the
\GEANTfour~\cite{GEANT4} program. Simulation and data events are
reconstructed with the same algorithms. The effect of pileup is simulated
in the MC samples in order to reproduce the observed pileup conditions in
data.

The signal samples for the pair production of top squarks ($\stp\stpb$)
are simulated for $250 \leq m(\stp) \leq 800$\GeV in steps of 25\GeV, and
$10 \leq$ \dm\ $\leq 80$\GeV in 10\GeV steps. The cross section for
$\stp\stpb$ production at NLO and including next-to-leading logarithmic
(NLL) corrections, as calculated by \PROSPINO v.2~\cite{prospino,Borschensky:2014cia,xs1,xs2,xs3,xs4,xs5},
varies approximately between 20 and 0.1 pb for the mass range considered.
The pair production of squarks with up to two additional jets from
initial-state radiation (ISR) is generated with \MGvATNLO~2.3.3 and is
then interfaced with \PYTHIA~8.212 for the decay, hadronization, and
showering. For the chargino-mediated decay of the scalar top, $m(\chg)$ is
taken to be the average of $m(\stp)$ and $m(\lsp)$. The decay is generated
to proceed via an off-shell \PW\ boson, and the \stp decay length is set
to zero. The modelling of the detector response is performed with the CMS
fast simulation program~\cite{Abdullin:2011zz}.

Simulated background and signal samples are corrected for differences with
respect to the values measured in data control samples in the selection
efficiencies for leptons and \cPqb\ jets, and for the misidentification
probability for light-quark and gluon jets as \cPqb\ jets. These
corrections are applied as functions of the \pt and $\eta$ of the objects.
For the signal samples, additional corrections are applied to take into
account any potential differences between the \GEANTfour and fast
simulations in regards to tagging efficiencies of \cPqb\ jets, leptons, and
modelling of \ptmiss.

\subsection{Preselection}
\label{s:presel}

The preselection requirements used in this paper are designed by
considering the general characteristics of the signal, and are based on
the methods presented in Ref.~\cite{Khachatryan:2015pot}. The CC and MVA
approaches share similar preselection requirements with a few minor
differences that are noted below due to studies showing that the MVA leads
to better performance with slightly different selection than the CC
search. In order to match the trigger requirement, events with $\ptmiss>200$
$(280)\GeV$ are selected for the CC (MVA) approach. This requirement
favours the signal, which tends to have larger missing transverse momentum
than SM processes due to two \lsp's escaping detection. The efficiency of
signal triggers is measured to be higher than 90 (98)\% for $\ptmiss>200$
$(280)\GeV$, and the simulated samples are reweighted as a function of
\ptmiss to account for the inefficiency.

Further suppression of SM processes such as \wjets\ is achieved by imposing
the additional requirement of $\HT>300\GeV$, where \HT is defined as the
scalar \pt sum of all jets. For the MVA search, this requirement is
relaxed to $\HT>200\GeV$. In order to improve the separation of signal and
SM background, we take advantage of events in which the \stp pair system
recoils against an ISR jet. In this case the LSP becomes Lorentz boosted,
which increases the \ptmiss in the event, while jets and leptons
remain relatively soft. The ISR jet candidate in the event is selected as
the leading jet with $\abs{\eta}<2.4$, which is required to satisfy $\pt>100$
$(110)\GeV$ for the CC (MVA) search. To reduce the contribution from
\ttbar production, events are required to have at most two jets with
$\pt>60\GeV$ in the CC search. In events with two jets, the azimuthal
angle between the leading and subleading (in \pt) jets is required to be
less than 2.5 radians in order to suppress the SM multijet background.

Finally, the soft single-lepton topology is selected by requiring at least
one muon or electron in the event, while vetoing events with a $\tau$
lepton, or a second electron or muon with $\pt>20\GeV$. At this stage of
the selection, the \wjets\ and \ttbar processes represent approximately
70\% and 20\% of the total expected background, respectively. The \zinv
process contributes with jets, genuine \ptmiss, and a jet misidentified as a
lepton. Diboson, single top and Drell--Yan (DY) processes also contribute,
with a lower expected yield due to a low cross section or a low acceptance
(or both).

\section{The CC approach}
\label{s:CCsearch}

\subsection{Signal selection}
\label{s:CCsel}

\begin{figure}
\centering
\includegraphics[width=.49\textwidth]{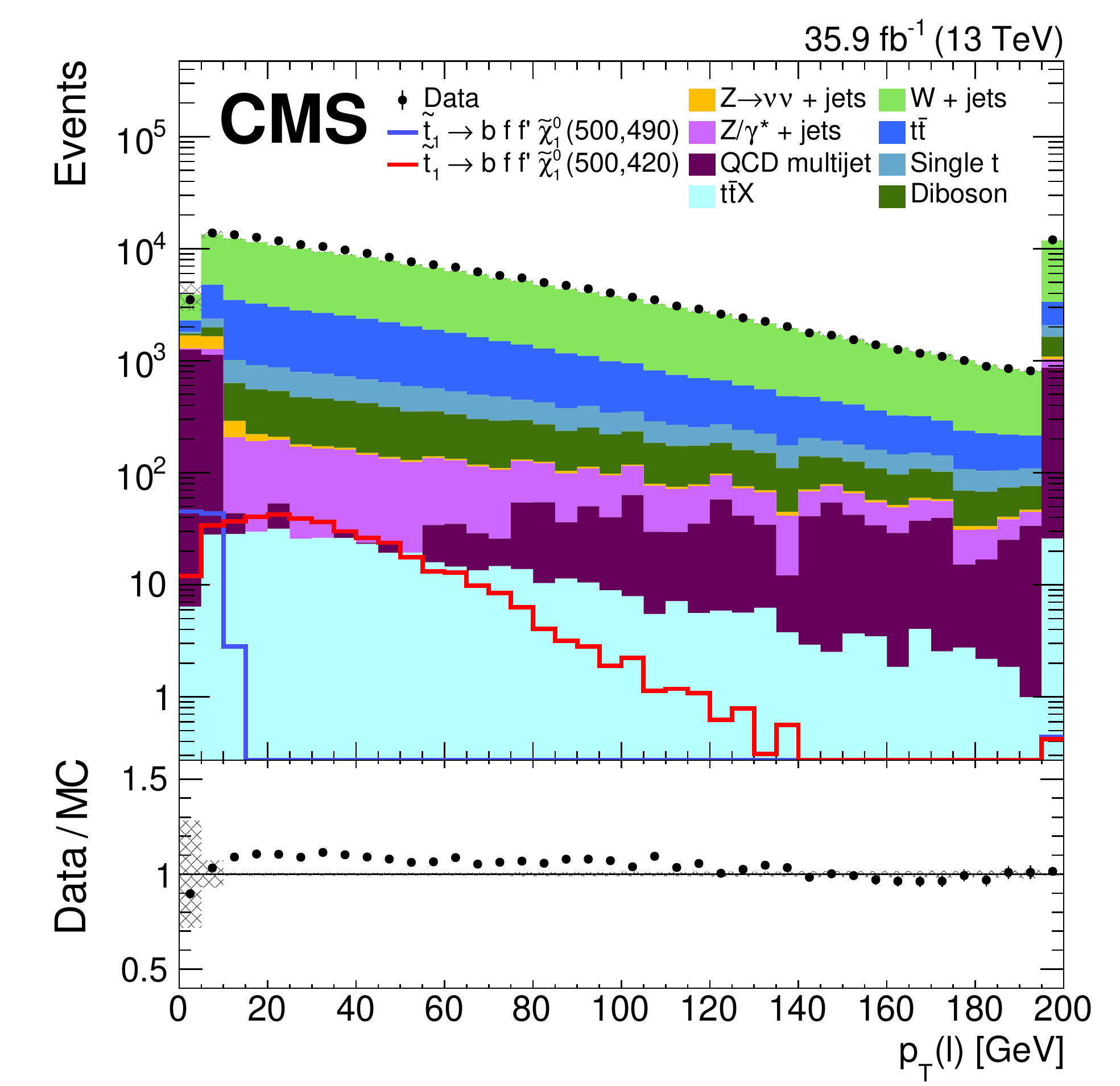} \hfil
\includegraphics[width=.49\textwidth]{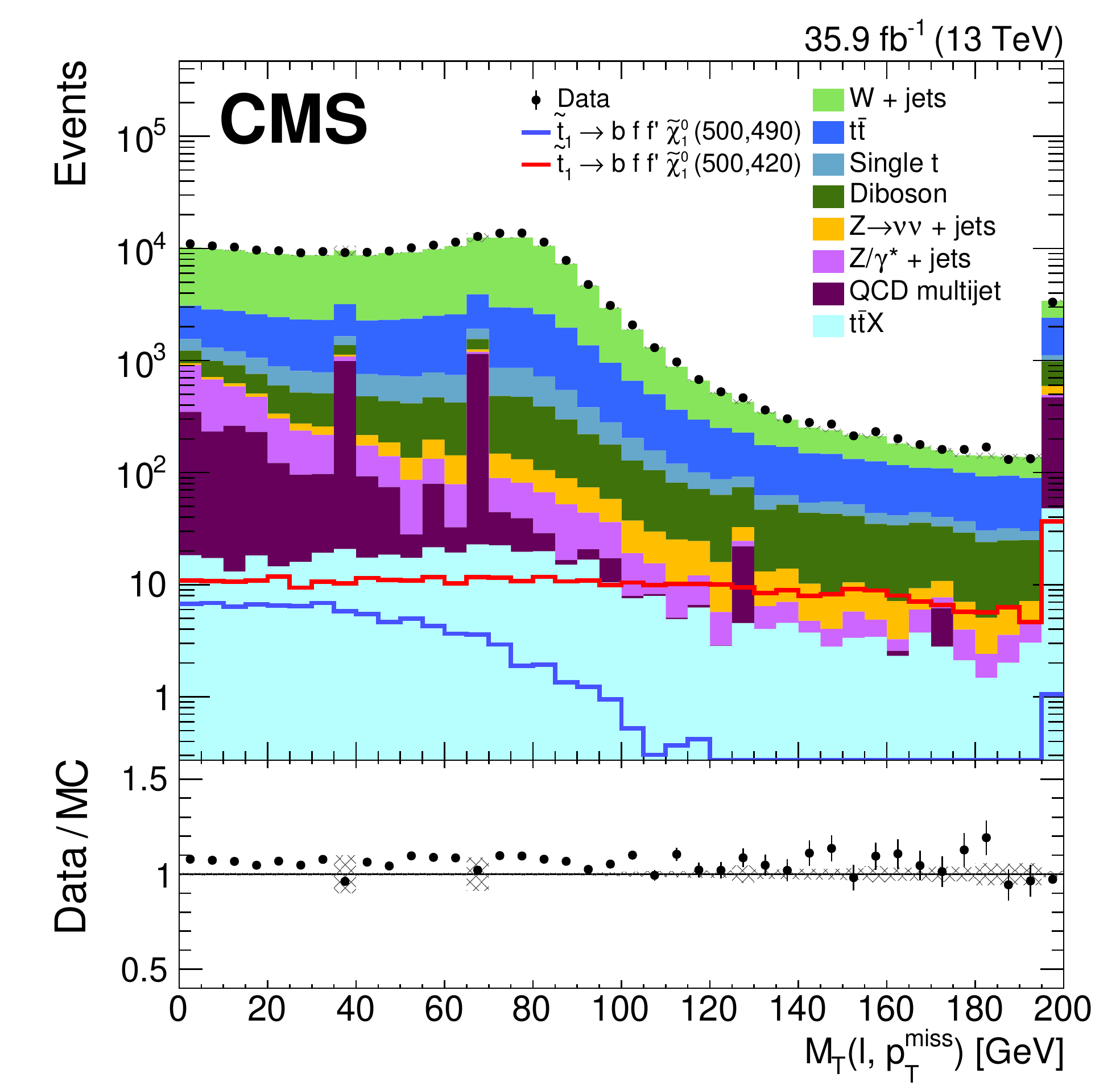}
\caption{Distributions of lepton \pt (left) and \mt~(right) at the
	preselection level in data and simulation. The lower panels show
	the ratio of data to the sum of the SM backgrounds where the dark
	shaded bands indicate the statistical uncertainties of the
	simulation. The distributions of two signal points of the
	four-body decay are also represented, while not being added to the
	background: \mmplane = (500,490) and (500,420)\GeV. The background
	distributions are obtained directly from simulation, and are
	normalized to an integrated luminosity of 35.9\fbinv. The last
	bin in each plot includes events beyond 200\GeV.
  }
\label{fig:CCpreselplots}
\end{figure}

After the preselection detailed in the previous section, \wjets\ is the
dominant background process, followed by a smaller contribution coming
from \ttbar production. A kinematic variable with good discrimination
against these background processes is the transverse mass $\mt \equiv
\sqrt{\smash[b]{2 \ptmiss \pt(\ell) (1-\cos \Delta\phi )}}$, where $\pt(\ell)$ is the
transverse momentum of the selected lepton and $\Delta\phi$ is the angular
difference between the lepton $\ptvec(\ell)$ and \ptvecmiss. The
distributions of lepton \pt and \mt~are shown in
Fig.~\ref{fig:CCpreselplots} for the observed data and simulated
background and signal, where we observe good agreement in the shapes of
the distributions between data and background simulation. The
normalization of the simulation is corrected by the results of a
background estimation technique based partially on data, as described in
Section~\ref{s:CCbkg}.

The signal regions (SRs) in the CC analysis are defined to maximize the
sensitivity of the search by exploiting the differences between the
kinematic properties of the final-state particles in the signal and
background processes. The leptons originating from the decay of the \stp
squark are expected to be much softer than those from SM processes.
Therefore, all SRs are required to satisfy $\pt(\ell)<30\GeV$. In order to
retain sensitivity to different \dm\ mass gaps, two signal regions SR1 and
SR2 are designed targeting small and large mass differences, respectively.
Moreover, due to the strong dependence of the \pt and \mt~distributions
for the signal on \dm, these SRs are further subdivided into a total of 44
mutually exclusive regions, which are detailed below and summarized in
Table~\ref{tab:CCsrcrdefs}. In this search, the correlations between \ptmiss,
\HT, and the transverse momentum of the ISR jet candidate
($\pt(\mathrm{ISR})$) are taken into account by defining SRs in terms of
the variables \ctone\xspace and \cttwo: $\ctone \equiv \min(\ptmiss, \HT -
100\GeV)$, and $\cttwo \equiv \min(\ptmiss, \pt(\mathrm{ISR}) - 25\GeV)$,
where the numerical values of 100 and 25\GeV are determined by
maximizing the ratio of signal to the square root of background in the
signal regions.

In SR1, events with a \cPqb\ jet satisfying $\pt>30\GeV$ are rejected since
the \cPqb\ jets in signal events with a small mass gap are expected to have
typical \pt values smaller than this threshold. This \cPqb-tag veto
significantly reduces the contribution of \ttbar events. In this region,
the \ptmiss and \HT requirements of the preselection are simultaneously
tightened by requiring $\ctone> 300\GeV$. Since the \wjets\ process is the
dominant background process for lower \mt~values, we take advantage of the
charge asymmetry in the production of W bosons at LHC and require the
lepton to have a negative charge in SR1 regions with $\mt<95\GeV$.
Moreover, the acceptance of the lepton is tightened by requiring
$\abs{\eta(\ell)}<1.5$, because leptons from decays of the W boson at the LHC
tend to be produced in the forward direction.

In SR2, we require at least one \cPqb\ jet with $\pt<60\GeV$, but reject
events with any \cPqb\ jet having $\pt>60\GeV$. These requirements increase
the efficiency of signal points with larger \dm\ while keeping the \ttbar
background under control. In this region we also require $\cttwo> 300\GeV$,
which is more effective in reducing the \ttbar background compared
to the \ctone\xspace requirement.

The SR1 (SR2) region is further divided in bins of \mt, lepton \pt, and
\ctone\xspace (\cttwo). The \mt~binning is done below and above the peak
around the W boson mass in the \mt~distribution, with the regions
$\mt<60\GeV$, $60<\mt<95$\GeV, and $\mt>95\GeV$ labelled as a, b, and c,
respectively. It can be seen from Fig.~\ref{fig:CCpreselplots} that the
lower (higher) \mt~region is more sensitive to signals with smaller
(larger) mass gaps. In order to take advantage of the shape differences in
lepton \pt distributions between various signal points and SM processes,
each SR is further divided into lepton \pt regions $5$--$12$, $12$--$20$,
and $20$--$30\GeV$, referred to as L, M, and H, respectively. An
additional region of $3.5$--$5.0$\GeV is added only for muons and only for
$\mt<95$\GeV, and is labelled VL. In addition, SR1 (SR2) is further
separated into two regions in \ctone\xspace (\cttwo) defined by $300 <
\ctone\xspace(\cttwo) < 400\GeV$ and $\ctone\xspace(\cttwo) > 400\GeV$
which are labelled X and Y, respectively.

\begin{table}
\centering
\topcaption{The CC search: definition of SRs. The subregions of SRs are denoted by
tags in parentheses, as described in the text: VL, L, M, and H refer to
the four bins in lepton \pt, and X and Y to the \ct~ranges specified in
the table. The corresponding control regions (CR) use the same selection
with the exception of the lepton \pt as shown in the table. For jets, the
attributes ``soft'' and ``hard'' refer to the \pt ranges $30$--$60\GeV$
and $>$60\GeV, respectively.}
\label{tab:CCsrcrdefs}
\cmsTable{
\begin{tabular}{l|c|c|c|c|c|c}
\hline
Variable                               & \multicolumn{6}{c}{Common to all SRs}       \\
\hline
Number of hard jets                    & \multicolumn{6}{c}{$\leq$2}               \\
$\Delta\phi$(hard jets) (rad)          & \multicolumn{6}{c}{$<$2.5}                \\
\ptmiss (\GeV)                         & \multicolumn{6}{c}{$>$300}     \\
Lepton rejection                       & \multicolumn{6}{c}{no $\tau$, or additional $\ell$ with $\pt>20\GeV$}           \\
\cline{2-7}
                                       & \multicolumn{3}{c|}{SR1}                & \multicolumn{3}{c}{SR2}           \\
\cline{2-7}
\HT (\GeV)                             & \multicolumn{3}{c|}{$>$400}                 & \multicolumn{3}{c}{$>$300}     \\
$\pt$(ISR jet) (\GeV)                  & \multicolumn{3}{c|}{$>$100}                 & \multicolumn{3}{c}{$>$325}     \\
Number of \cPqb\ jets                  & \multicolumn{3}{c|}{$0$}                    & \multicolumn{3}{c}{$\geq$1 soft, $0$ hard} \\
$|\eta(\ell)|$                            & \multicolumn{3}{c|}{$<$1.5}                 & \multicolumn{3}{c}{$<$2.4}     \\
\cline{2-7}
                                       & SR1a     & SR1b              & SR1c                 & SR2a   & SR2b       & SR2c           \\
\cline{2-7}
\mt (\GeV)                             & $<$60    & $60$--$95$        & $>$95        & $<$60    & $60$--$95$        & $>$95        \\
$Q(\ell)$                                 & $-1$     & $-1$ &            any          & any & any & any        \\
$\pt(\mu)$ (\GeV)                      & $3.5$--$5$ (VL)      & $3.5$--$5$ (VL)      &                  \NA
                                       & $3.5$--$5$ (VL)      & $3.5$--$5$ (VL)      &                  \NA   \\
$\pt(\Pe,\mu)$ (\GeV)                  & $5$--$12$ (L)          & $5$--$12$ (L)          & $5$--$12$ (L)
                                       & $5$--$12$ (L)          & $5$--$12$ (L)          & $5$--$12$ (L)          \\
                                       & $12$--$20$ (M)         & $12$--$20$ (M)         & $12$--$20$ (M)
                                       & $12$--$20$ (M)         & $12$--$20$ (M)         & $12$--$20$ (M)         \\
                                       & $20$--$30$ (H)         & $20$--$30$ (H)         & $20$--$30$ (H)
                                       & $20$--$30$ (H)         & $20$--$30$ (H)         & $20$--$30$ (H)         \\
                                       & $>$30 (CR)         & $>$30 (CR)         & $>$30 (CR)
                                       & $>$30 (CR)         & $>$30 (CR)         & $>$30 (CR)         \\
\hline

\CT (\GeV)                             & \multicolumn{3}{c|}{$300<\ctone<400$ (X)}
                                       & \multicolumn{3}{c}{$300<\cttwo<400$ (X)}                             \\
                                       & \multicolumn{3}{c|}{$\ctone>400$ (Y)}
                                       & \multicolumn{3}{c}{$\cttwo>400$ (Y)}                               \\

\hline
\end{tabular}
}
\end{table}

\subsection{Background prediction}
\label{s:CCbkg}

The dominant backgrounds in most of the CC signal regions are \wjets\ and
\ttbar production with a prompt lepton in the final state. The nonprompt
sources of leptons become more important in regions with large \mt~or very
low lepton \pt. In this section, the methods used to estimate the prompt
and nonprompt backgrounds from data are described. Simulation is used to
estimate other rare backgrounds with a prompt lepton, namely \Zg, diboson,
single top quark production, and \ttbar production with an additional \PW,
\Z, or $\gamma$.

The nonprompt background due to misidentified leptons associated with a
jet becomes comparable to the prompt contribution in regions where
\wjets\ and \ttbar production are suppressed, namely in regions of
high-\mt~and very low lepton \pt. This background is estimated fully from
data using the ``tight-to-loose'' method, where a ``loose'' set of
identification and isolation criteria are defined to select lepton
candidates that are more likely to be nonprompt. The loose selection is
defined by relaxing the requirement on the lepton isolation to
$\Iabs<20\GeV$ for $\pt(\ell)<25\GeV$ and $\Irel<0.8$ for
$\pt(\ell)>25\GeV$, as well as relaxing the impact parameter conditions to
$\abs{\dxy}<0.1\unit{cm}$ and $\abs{\dz}<0.5\unit{cm}$. The ``tight'' criteria
correspond to the final lepton selection of the analysis, described in
Section~\ref{s:objects}. The probability that a loose lepton also passes
the tight criteria, the tight-to-loose fraction $\epsilon_\mathrm{TL}$, is
measured as a function of lepton \pt and $\abs{\eta}$ in an orthogonal
``measurement region'' largely dominated by multijet events, which is
enriched in nonprompt leptons. The fraction $\epsilon_\mathrm{TL}$ is
measured from data, after the subtraction of the simulated prompt lepton
contribution. The final estimate of nonprompt leptons in a SR or control
region (CR) is based on the observed yield in an "application region". The
latter is defined in the same way as the corresponding SR or CR, with the
exception that the lepton has to pass the loose lepton criteria but not
the tight ones. The final estimate is obtained by scaling the data yield
in the application region by
$\epsilon_\mathrm{TL} / (1-\epsilon_\mathrm{TL})$, after subtracting
the simulated prompt lepton contribution.

The absolute normalization of the prompt background simulation in each SR
is obtained from a CR with identical requirements as in the SR except for
the lepton \pt selection. The CR is defined by replacing the lepton \pt
requirement of the SR with $\pt(\ell)>30\GeV$; therefore, SRs that are
only distinguished by different selections in $\pt(\ell)$ share the same
CR. The impact of potential signal contamination is taken into account
when deriving the results as described in Section~\ref{s:res}. In each CR,
a scale factor for the prompt simulation is obtained by normalizing the
simulation to data, after subtracting nonprompt and rare background
sources from the observed number of events in the CR. The nonprompt
contribution used in the subtraction is estimated separately from data.
The composition of the CRs in terms of background processes, as well as
the total simulated and observed yields, are shown in
Table~\ref{tab:CCcrbkg}. The scale factors, ranging from 0.86 to 1.25, are
then applied to the simulation in the corresponding SRs. In order to
verify the extrapolation of the scale factors from CR to SR, we perform
the same background estimation procedure in validation regions (VRs),
which are orthogonal to all SRs and CRs. Each validation region is
obtained by one of the following changes: (a) lowering the \ctone~(in SR1
and CR1) and \cttwo~(in SR2 and CR2) requirements to $200 < \CT <300\GeV$,
(b) replacing the conditions on \cPqb\ jets by requiring at least one
\cPqb\ jet with $\pt>60\GeV$. The predictions in the validation regions are
compatible with the observations within the uncertainties.

\begin{table}
\centering
\topcaption{The CC search:
    observed yields and simulated background contributions to CRs
    normalized to an integrated luminosity of 35.9\fbinv.
    The nonprompt contributions are estimated from data.
    The last column shows the scale factors used for the normalization of the
    \wjets\ and \ttbar samples.
    Only statistical uncertainties are reported.
}
\label{tab:CCcrbkg}
\cmsTable{
\begin{tabular}{lccccccc}
\hline
Region     & \wjets             & \ttbar & Nonprompt & Rare & Total SM & Data     &  Scale factors                \\
\hline
CR1aX &   $2133\pm20$   & $226.6\pm3.5$  & $44.5\pm6.4$  & $293.2\pm5.9$  & $2698\pm22$  & 2945   &  $1.10\pm0.03$\\
CR1aY &   $878.3\pm8.6$ & $65.8\pm1.9$  & $13.3\pm3.6$  & $139.4\pm4.1$  & $1097\pm10$  & 1197   &  $1.11\pm0.04$\\
CR1bX &   $1107\pm15$   & $134.5\pm2.7$  & $7.8\pm2.7$  & $112.1\pm4.1$  & $1361\pm16$  & 1462   &  $1.08\pm0.03$\\
CR1bY &   $438.2\pm6.4$  & $35.1\pm1.4$  & $1.6\pm1.6$  & $51.9\pm2.9$ & $526.8\pm7.3$ & 502    &  $0.95\pm0.05$\\
CR1cX &   $642\pm11$   & $103.8\pm2.3$  & $12.7\pm3.0$  & $174.3\pm5.5$  & $932\pm13$  & 1051   &  $1.16\pm0.05$\\
CR1cY &   $278.3\pm8.3$  & $25.5\pm1.2$  & $6.2\pm2.2$  & $102.2\pm4.3$  & $412.2\pm9.6$ & 432    &  $1.07\pm0.08$\\
CR2aX &   $171.7\pm2.5$  & $195.6\pm3.3$  & $1.9\pm1.9$  & $64.2\pm1.9$  & $433.4\pm4.9$ & 451    &  $1.05\pm0.06$\\
CR2aY &   $74.5\pm1.0$ & $58.4\pm1.7$  & $0.8\pm0.8$ & $25.6\pm1.1$  & $159.3\pm2.4$ & 145    &  $0.89\pm0.09$\\
CR2bX &   $104.9\pm2.0$  & $110.8\pm2.5$  & $1.2\pm1.2$  & $39.2\pm1.6$  & $256.1\pm3.8$ & 226    &  $0.86\pm0.07$\\
CR2bY &   $42.6\pm0.8$ & $30.8\pm1.3$  & $0.3\pm0.3$ & $15.0\pm0.9$ & $88.6\pm1.8$ & 79     &  $0.87\pm0.12$\\
CR2cX &   $17.3\pm0.8$ & $53.8\pm1.7$  & $1.7\pm1.2$  & $15.7\pm1.0$  & $88.4\pm2.4$ & 106    &  $1.25\pm0.15$\\
CR2cY &   $7.5\pm0.8$ & $12.8\pm0.8$ & $0.6\pm0.6$ & $6.6\pm0.7$ & $27.5\pm1.5$ & 29     &  $1.07\pm0.28$\\
\hline
\end{tabular}
}
\end{table}

\subsection{Systematic uncertainties}
\label{s:CCsys}

Processes for which the absolute yield is predicted by simulation are
subject to systematic uncertainties in the determination of the integrated
luminosity (2.5\%)~\cite{CMS-PAS-LUM-17-001}. All simulated samples are
subject to experimental uncertainties on the jet energy scale (JES) and
jet energy resolution (JER). The uncertainties due to miscalibration of
the JES are estimated by varying the jet energy corrections up and down by
one standard deviation and propagating the effect to the calculation of
\ptmiss. Moreover, differences of the JER between data and simulation are
accounted for by smearing the momenta of jets in simulation. The
uncertainties corresponding to \cPqb-tagging efficiencies and
misidentification rates for tagging light-flavoured or gluon jets as
\cPqb\ jets have been evaluated for all simulated samples. The
uncertainties corresponding to the correction of simulated samples for
trigger and lepton efficiencies are taken as systematic uncertainties. The
uncertainty due to the simulation of pileup for simulated background
processes is taken into account by varying the expected cross section of
inelastic collisions by 5\%~\cite{Sirunyan:2018nqx}. An uncertainty of
50\% is assigned to the cross sections of all nonleading backgrounds. An
overview of all systematic uncertainties related to the background
prediction is presented in Table~\ref{tab:CCbkgsys}.

The nonprompt background estimation method of this search, as described in
the previous section, depends on the tight-to-loose fraction
$\epsilon_\mathrm{TL}$ which is sensitive to the flavour content of jets.
The systematic uncertainty due to possible differences in the flavour
content of jets between the measurement and application regions is
assessed by varying the \cPqb-tagging requirements of the measurement
region. The resulting uncertainty ranges from 20 to 50\% from low to high
lepton \pt, respectively. The consistency of the method is tested by
applying the same procedure to simulated data. To account for any residual
deviation found in the test, an additional uncertainty of 20 to 200\% is
assigned in some regions, with the highest uncertainties applying to
regions that are dominated by prompt background.

The prompt background prediction procedure of this search, as described in
the previous section, relies on the simulation of \wjets\ and \ttbar
production and is sensitive to theoretical uncertainties on ISR. The
modelling of ISR for these processes is checked in control samples in data
that are highly enriched in \ttbar or \wjets\ events. The simulation of
\ttbar events is tested by comparing the jet multiplicity observed in a
control sample with the simulation. Simulated \ttbar events are reweighted
based on this comparison, and half of the correction is assigned as the
systematic uncertainty~\cite{Sirunyan:2017cwe}. This systematic
uncertainty affecting \ttbar also affects the signal samples. Similarly,
the simulation of \wjets\ events is corrected based on the distribution of
$\pt(\PW)$ in a control sample, and the difference between the uncorrected
and the corrected simulation is assigned as a systematic
uncertainty~\cite{Sirunyan:2017zss}. These two sources of uncertainties
lead to relative changes of the total background estimation in the SRs
that range from 2 to 10\% for the \wjets\ process, and are less than 1\%
for the \ttbar process.  The estimate of the prompt background depends
only weakly on the background composition, since the distributions of
$\pt(\ell)$ in \wjets\ and \ttbar processes are similar. The corresponding
systematic uncertainty is derived from a 20\% variation in the relative
yields of \wjets\ and \ttbar backgrounds.

The dominant source of systematic uncertainty for the signal is caused by
the modelling of ISR. It is minimized by reweighting the jet multiplicity
in the signal sample according to the corrections obtained in the \ttbar
sample. Uncertainties due to unknown higher-order effects are estimated by
variations of the renormalization and factorization scales by factors of
0.5 and 2~\cite{syscalc}. Moreover, possible differences between the fast and the full
\GEANTfour-based modellings of \ptmiss are taken into account and the
corresponding uncertainties are assigned to the signal yields as shown in
Table~\ref{tab:CCbkgsys}. The statistical uncertainty of the signal
simulation ranges from 8 to 15\%.

\begin{table}
\centering
\topcaption{The CC search: typical ranges for
relative systematic uncertainties (in \%) on the
  total background prediction and signal prediction in the main SRs.
  The ``\NA'' means that a certain source of uncertainty is not applicable.}
\label{tab:CCbkgsys}
\begin{tabular}{l c c c }
\hline
Systematic                                  & \multicolumn{2}{c}{Background} & Signal \\
uncertainty                                 & SR1 & SR2                       &        \\
\hline
Renormalization \&            &        &         &     \\
factorization scales              & \NA       & \NA         &  2--3   \\
Pileup                            & 0.1--1.8 & 0.1--2.0   &  1     \\
JES                               & 1.2--2.1 & 0.1--1.4   &  3--4   \\
JER                               & 0.1--0.5 & 0.1--1.1   &  0--1   \\
\cPqb-tagging                     & 0.1     & 0.1--1.0   &  1--3   \\
Trigger                           & 0.0--0.1  & 0.0--0.1     & 1     \\
Lepton efficiency                 & 1.0--1.8 & 1.0--1.5   &  3     \\
ISR (\ttbar and signal)           & 0.1--0.5 & 0.1--0.8   &  5--7     \\
ISR (\wjets)                      & 4.5--10.2& 1.9--4.4   &  \NA     \\
\ptmiss modelling ({\sc FastSim})     & \NA       & \NA          & 2--3   \\
Relative yields \wjets/\ttbar     & 0.1--1.6 & 0.1--2.2   &  \NA     \\
Nonprompt                         & 1.0--4.6   & 1.0--9.5      & \NA     \\
\hline
\end{tabular}
\end{table}

\section{The MVA approach}
\label{s:MVAsearch}

\subsection{Signal selection}
\label{s:MVAsel}

For the selection of the signal events corresponding to four-body decays
of the \stp, we use a boosted decision tree
(BDT)~\cite{RefBDT,Hocker:2007ht} to take advantage of the correlations
among variables that discriminate between signal and background.

Compared to the approach of Ref.~\cite{Khachatryan:2015pot}, we use new
variables and search for the most reduced set of best-performing variables
to be used as input to the BDT. To find the most discriminating variables
we test different sets maximizing the figure of merit (FOM)~\cite{cowan}:
\begin{equation}
\mathrm{FOM} = \sqrt{ 2 \left( (S+B)\ln\left[\frac{(S+B) \ (B+\sigma_B^2)}{B^2 + (S+B) \ \sigma_B^2}\right] - \left(\frac{B^2}{\sigma_B^2}\right)\ln\left[1 + \frac{\sigma_B^2 \ S}{B \ (B+\sigma_B^2)}\right] \right) },
\label{eq:fom}
\end{equation}
where S and B stand for the expected signal and background yields. The
term $\sigma_B = (f \, B)$ represents the expected systematic
uncertainty on the background with $f$ being an estimate of the relative
uncertainty of the background yield, taken to be $f=20\%$ (see
Section~\ref{s:MVAsys}). A new variable is incorporated into the set of
input variables only if it significantly increases the FOM. The full list
of the final input variables is:
\begin{itemize}
\item \ptmiss, \ptl, and \mt: The correlation between \ptmiss and \ptl differs
  between signal, where the \ptmiss is due to three missing objects (two \PSGczDo
  and a $\nu$), and \ttbar and \wjets\ backgrounds where the \ptmiss is due to
  a single missing object ($\nu$). The \mt~distribution peaks at $\approx$80\GeV
  for SM processes where a W boson is produced, while being a
  rather broad distribution for signal.
\item \etl and \chgl: The pseudorapidity of the lepton \etl is considered because
  the decay products of the signal are more centrally produced than those of the
  \wjets\ background. The charge of the lepton \chgl~is also considered,
  as \PWp and \PWm are produced unequally at the LHC, while
  the signal events contain equal numbers of positive and negative leptons.
\item \ptisr, \ptb, \njets, and \HT:
  The \pt of the leading jet, \ptisr, captures the hard ISR jet in signal events,
  and \ptb\ is the \pt of the \cPqb\ jet with the highest \cPqb\
tagging discriminant value. Both are sensitive to
  the different phase space available for signal and background events:
  $m(\cPqt)-m(\PW)$ for \ttbar, and $m(\stp)-m(\lsp)$ for signal.
  The multiplicity of selected jets \njets~is included, reflecting
  the mass of the mother particle \stp.
\item \nbl, \drLB, and \bdisc: The number of loose \cPqb\ jets \nbl, the
  distance between the lepton and the jet with the highest \cPqb\ tagging discriminant
  \drLB, and the highest \cPqb\ tagging discriminant per event \bdisc~are
  also included among the input variables.
\end{itemize}

\begin{figure}[!htbp]
\centering
\includegraphics[scale=0.39]{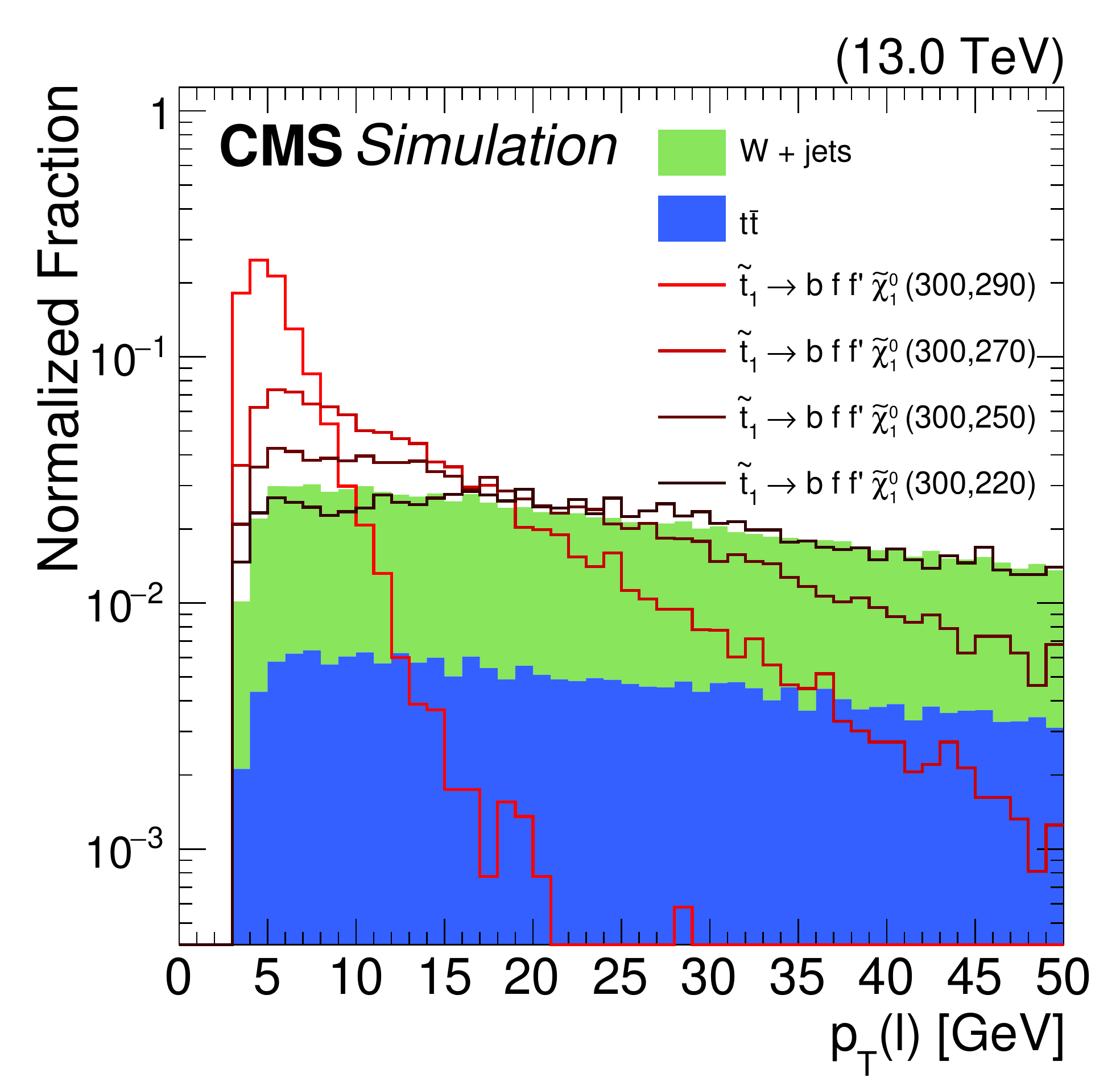}
\includegraphics[scale=0.39]{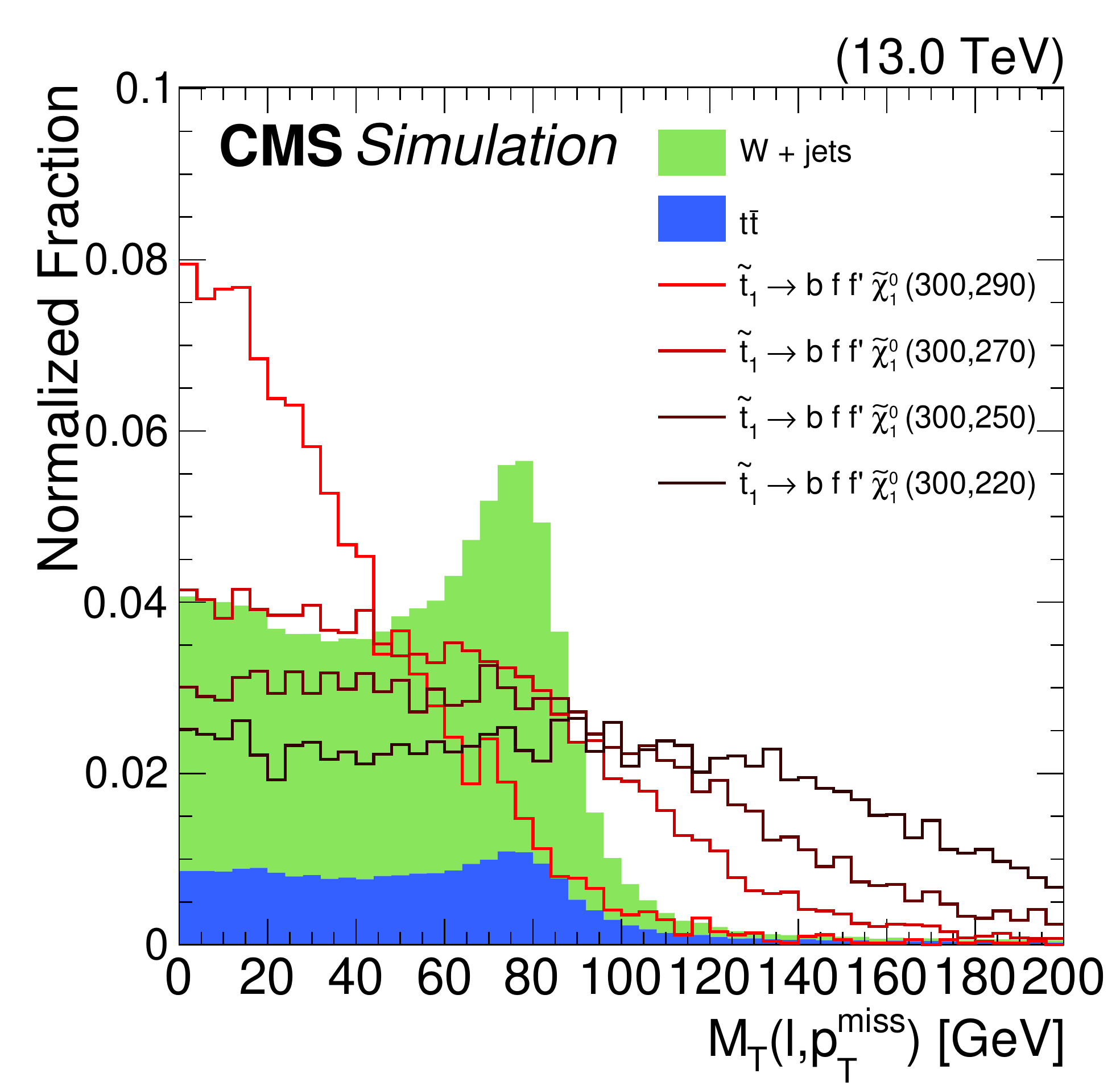}
\caption{Simulated distributions of \ptl (left) and \mt~(right) at the
  preselection level for signal samples with different \dm, and \wjets\ and \ttbar
  background events. The area of each signal distribution, and the total
  background contribution, are normalized to unit area.
}
\label{fig:vardm}
\end{figure}

Because the discrimination power of each input variable varies as a
function of \dm, as shown in Fig.~\ref{fig:vardm}, the \mmplane plane is
partitioned into eight \dm\ regions (from 10 to 80\GeV, in 10\GeV steps)
and a separate BDT is trained for each partition. The \wjets\ and \ttbar
processes, which represent a large fraction of the total background after
preselection, are included in the training of the BDT. The \zinv process,
which represents a nonnegligible fraction of the remaining total
background at the final selection level, is also included. The training is
done with simulated events for both signal and background processes. The
background samples are normalized to their respective cross section to
realistically represent the SM background in the training. We take
advantage of the similar distribution of the input variables for different
\mmplane signal points with the same \dm, and regroup all signal points
for a given \dm\ together when feeding signal to the BDT training. This
increases the number of signal events for each training. Due to the large
variation of the spectrum of the \ptl variable across the \mmplane plane,
we require $\ptl < 30$\GeV for signal points with $\dm \leq 60$\GeV
before training different BDTs, while there is no restriction on \ptl for
signal points with higher \dm.

\begin{figure}[!htbp]
\centering
\includegraphics[width=0.49\textwidth]{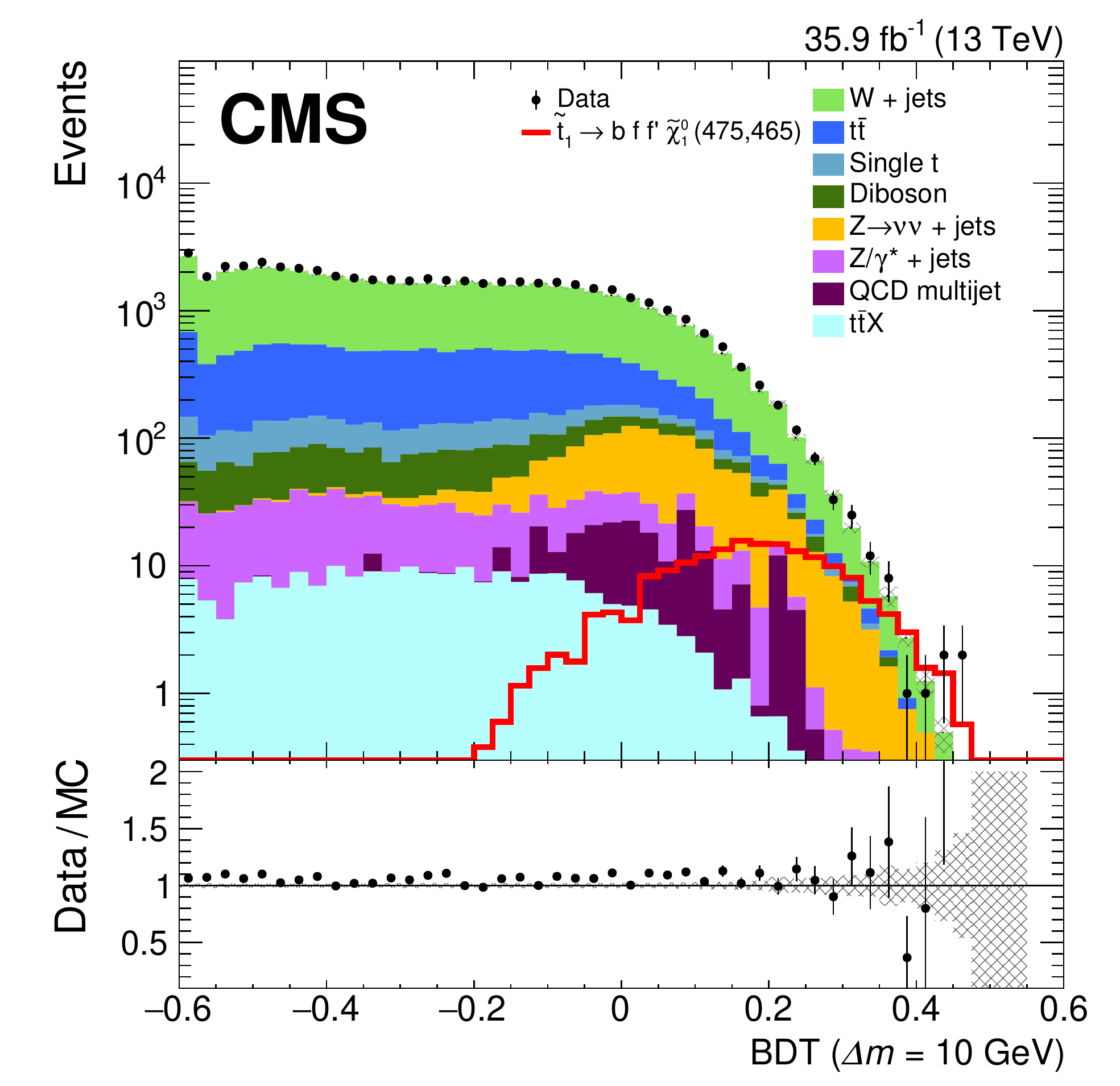} \hfil
\includegraphics[width=0.49\textwidth]{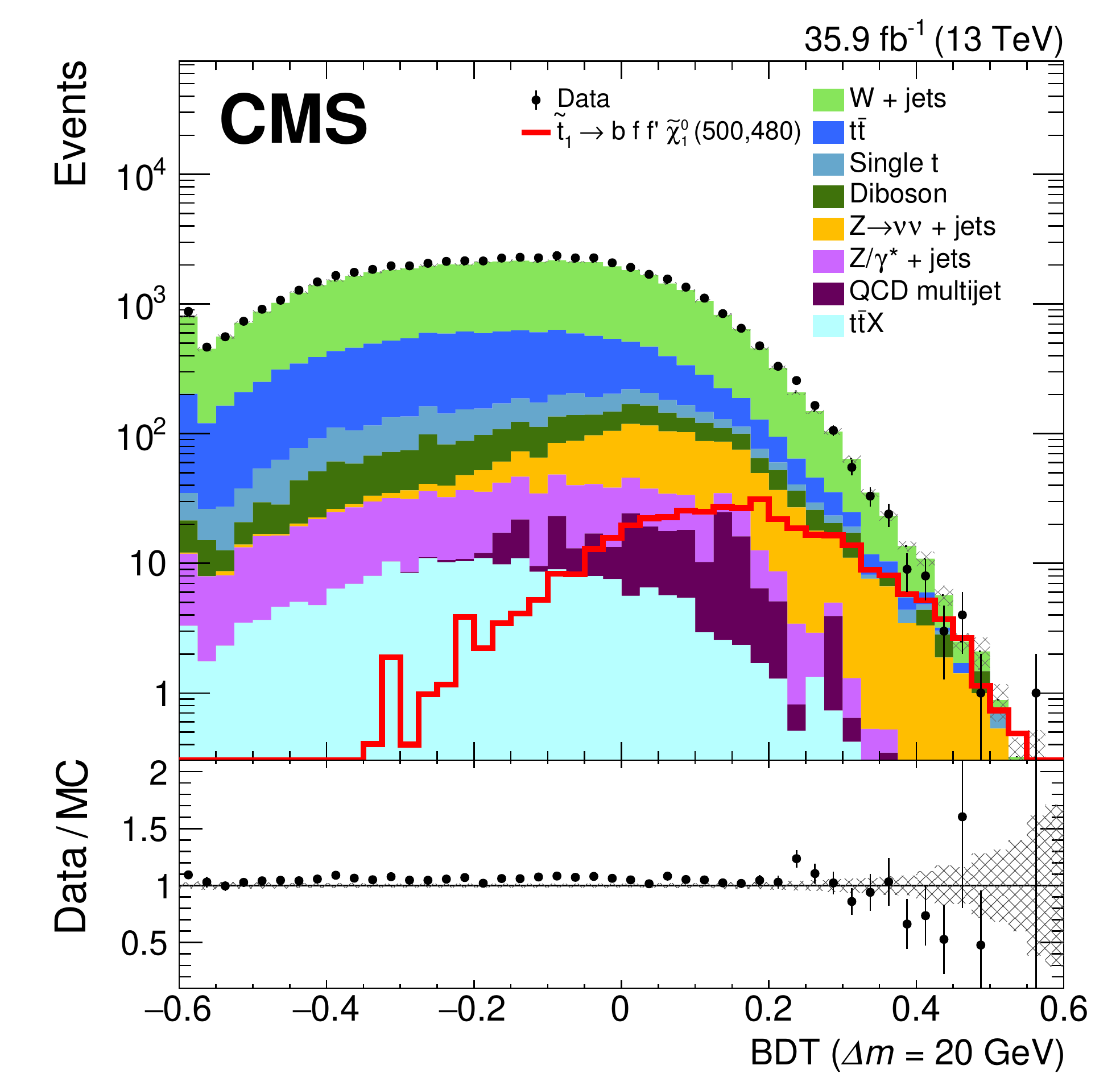} \\
\includegraphics[width=0.49\textwidth]{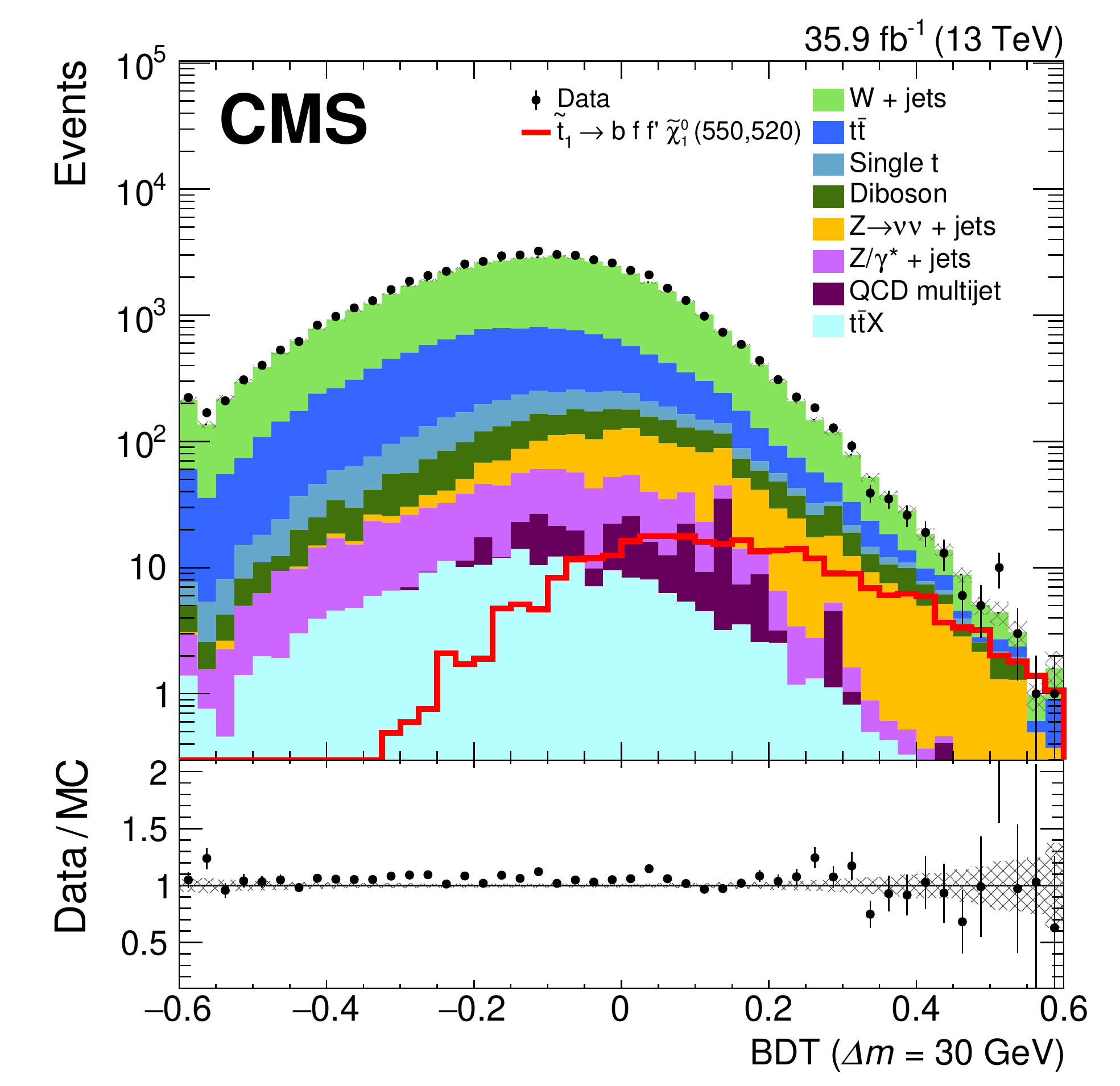} \hfil
\includegraphics[width=0.49\textwidth]{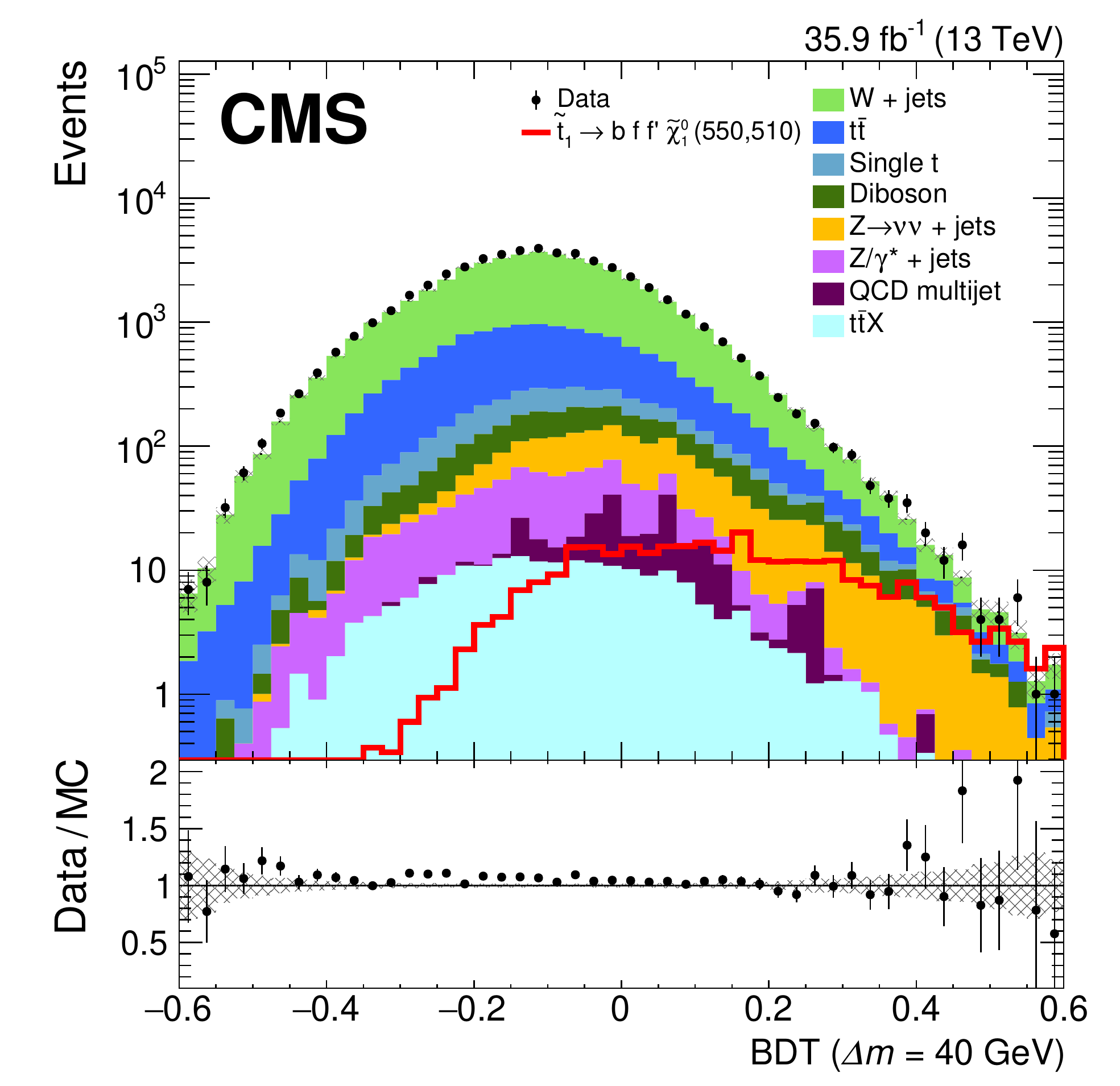} \hfil
\caption{Distributions of the BDT output at the preselection level in data and
        simulation in 10\GeV steps of \dm\ from 10 (top-left) to 40\GeV
        (bottom-right). For each case, a representative \mmplane signal point
        is also shown, but is not added to the SM background. The shaded area
        on the Data/MC ratio represents the statistical uncertainty of the
        simulated background.
}
\label{fig:bdtdatamc1}
\end{figure}

\begin{figure}[!htbp]
\centering
\includegraphics[width=0.49\textwidth]{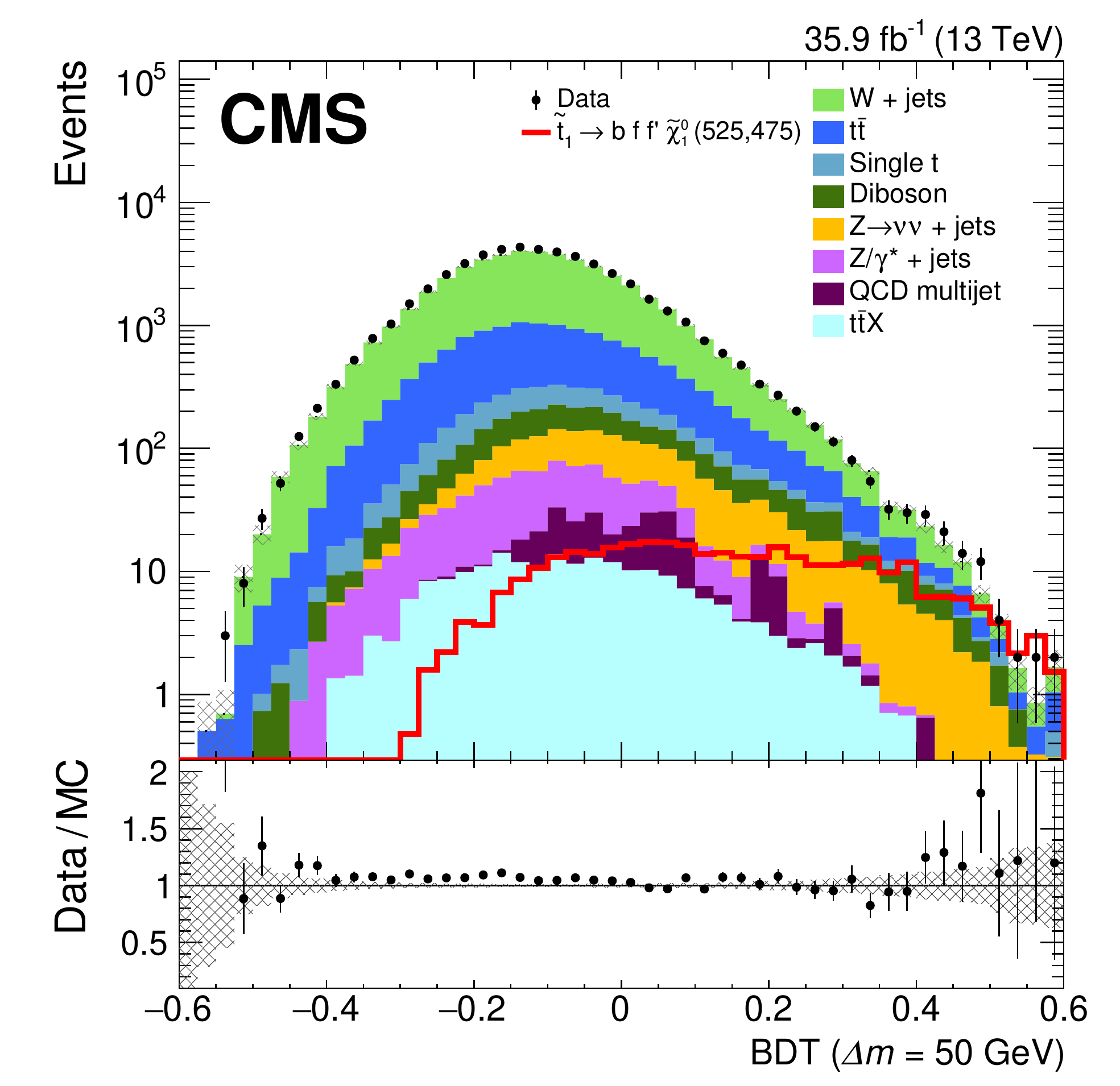} \hfil
\includegraphics[width=0.49\textwidth]{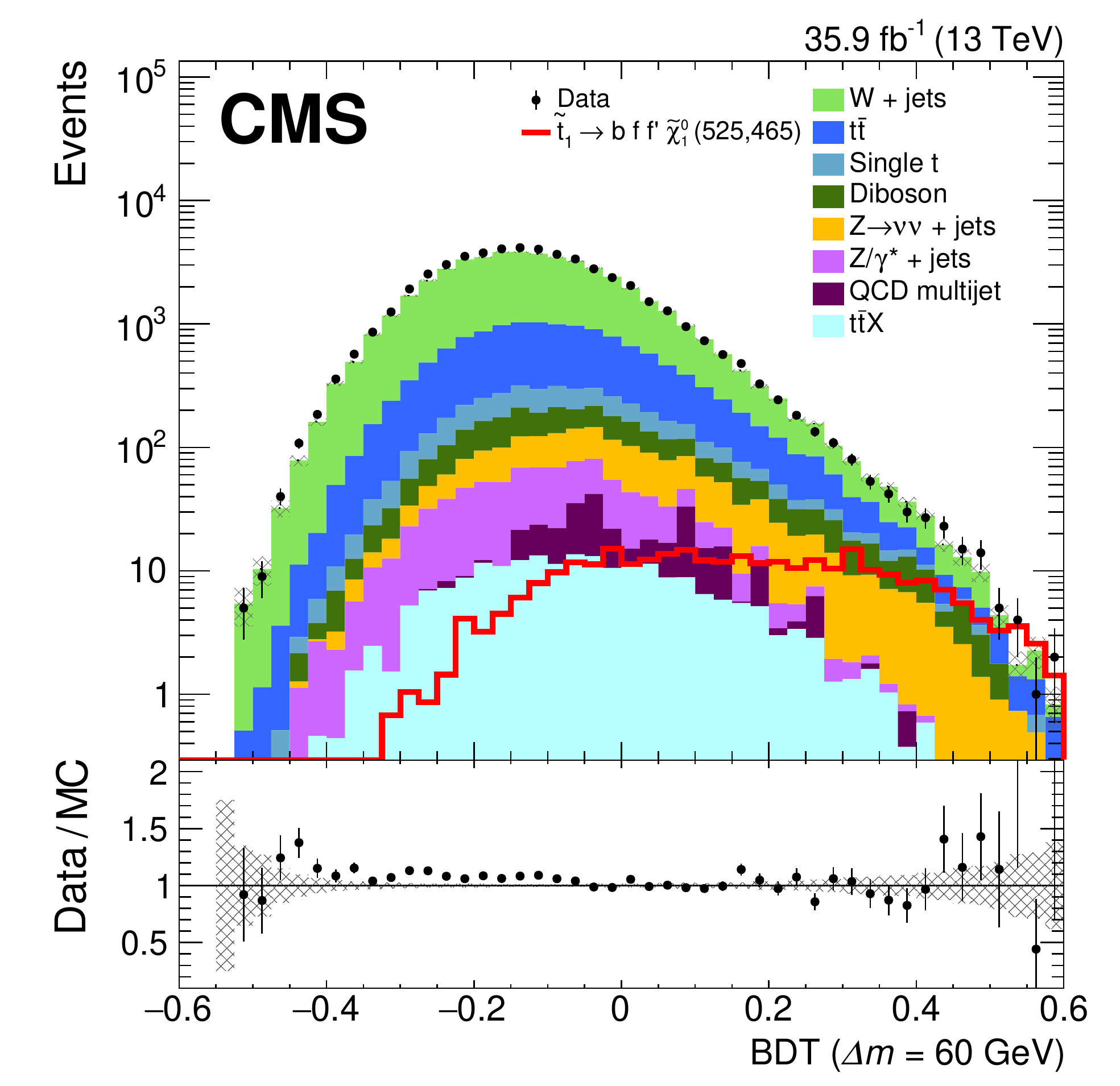} \\
\includegraphics[width=0.49\textwidth]{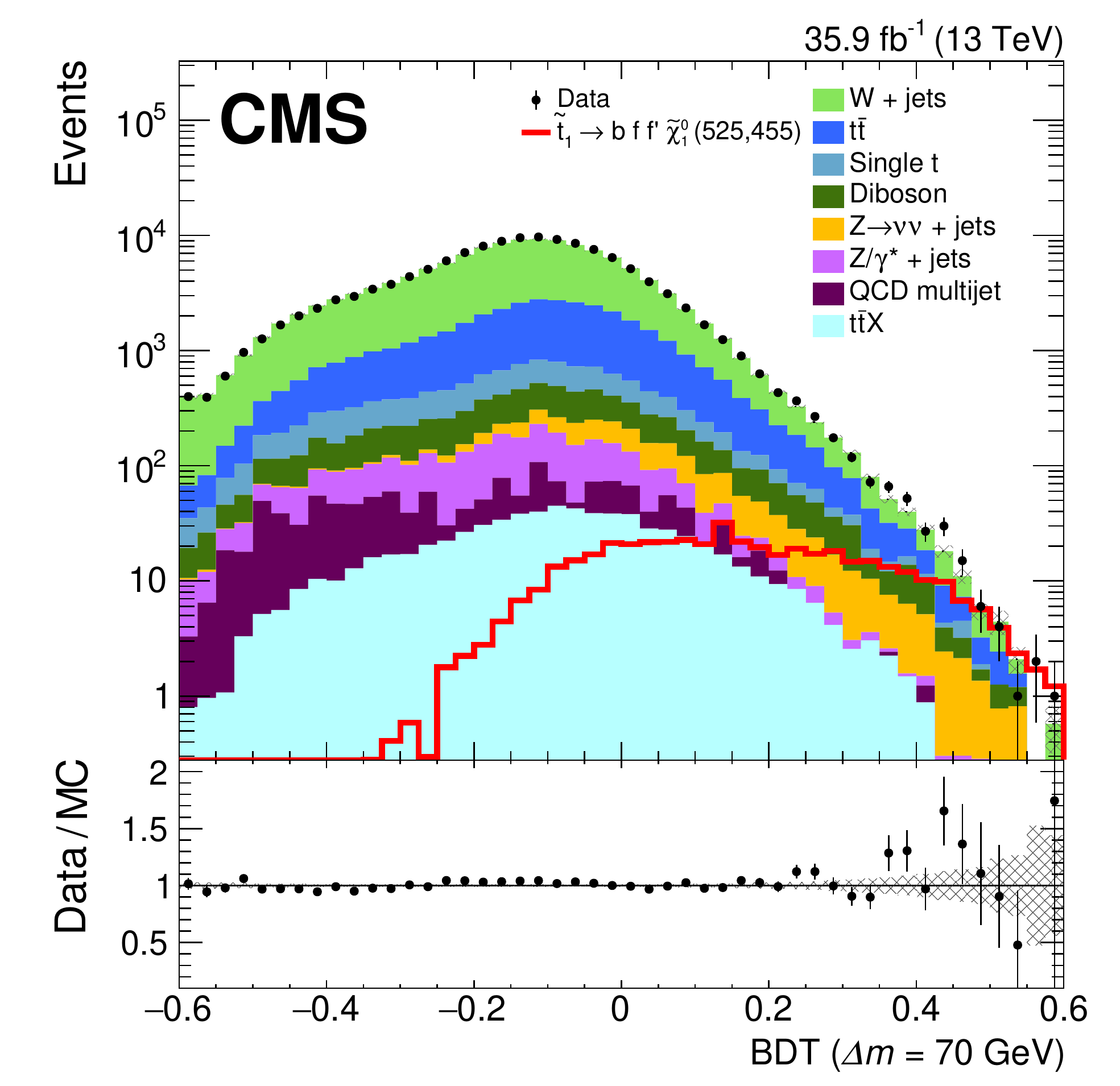} \hfil
\includegraphics[width=0.49\textwidth]{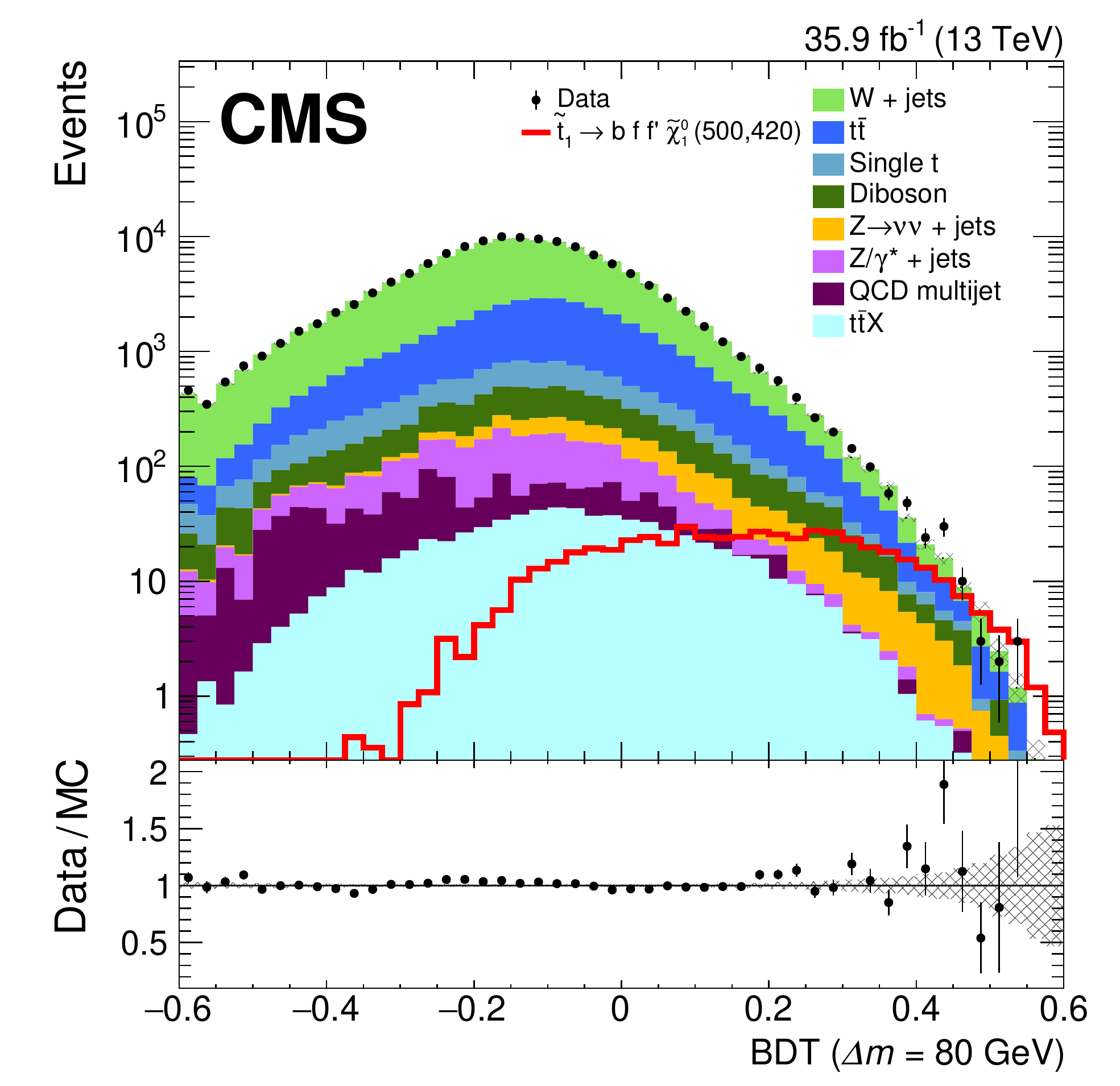}
\caption{Distributions of the BDT output at the preselection level in data and
        simulation in 10\GeV steps of \dm\ from 50 (top-left) to 80\GeV
        (bottom-right). For each case, a representative \mmplane signal point
        is also shown, but is not added to the SM background. The shaded area
        on the Data/MC ratio represents the statistical uncertainty of the
        simulated background.
}
\label{fig:bdtdatamc2}
\end{figure}

Figures~\ref{fig:bdtdatamc1} and~\ref{fig:bdtdatamc2} show the output
distribution of the BDT in data and for the total SM background as taken
from simulation. In each case a representative \mmplane signal point is
also shown, chosen at the limit of the expected sensitivity of the CC
search (see Section~\ref{s:res}) and belonging to the \dm\ for which the
training has been done. We observe that the responses of the BDT,
henceforth called BDT outputs, are not the same. This is due to the
changing mix between signal and background as well as varying differences
of correlations across the \mmplane plane, resulting in different BDT
outputs for different \dm\ values. We observe good agreement between data
and simulation over the entire range dominated by the background (\eg BDT
output smaller than 0.3) for the eight different trainings. The BDT output
is also checked in data to be well reproduced by the simulation in two
validation regions, across the entire range of the BDT output. These
regions are kinematically orthogonal to the preselection while using the
same online selection, and are defined as follows:
\begin{itemize}
\item Preselection with $200 < \ptmiss < 280\GeV$,
\item Preselection with $\ptl > 30$\GeV.
\end{itemize}
They are also used to evaluate the precision of the method for predicting
background, as described in Section~\ref{s:MVAbckg}. A SR is defined by
applying a threshold to each BDT output. The thresholds on the BDT output
are reported in Table~\ref{tab:mvares}. On average the BDT selection
suppresses the SM background by a factor ${\approx}3\times10^3$ while
reducing the signal by a factor $\approx$25. The total efficiency for
signal points at the limit of the sensitivity of the CC search, and across
all selections, is of the order of $1.3\times10^{-4}$.

\subsection{Background predictions}
\label{s:MVAbckg}

The predicted numbers of \wjets\ and \ttbar events are obtained from data
control regions (CRs) based on the BDT output. The number of estimated
prompt background events in the SR, $Y^\text{SR}_\text{prompt}$, is
derived as follows:
\begin{equation}
\begin{aligned}
Y^\text{SR}_\text{prompt}(X) & = & N^\text{SR}_\text{prompt}(X) \ \left[\frac{N^\text{CR}(\text{Data}) - N^\text{CR}_\text{prompt}(\text{Rare}) - N^\text{CR}_\text{nonprompt}}{N^\text{CR}_\text{prompt}(X)} \right].
\label{eq:ddbckg}
\end{aligned}
\end{equation}
Here, $X$ refers to background processes to be estimated, \wjets\ or
\ttbar. The superscripts SR and CR, respectively, refer to the signal and
control regions. The term ``prompt'' refers to processes where a prompt
lepton is produced. The term ``nonprompt'' refers to processes where there
is a jet misreconstructed as a lepton. The numbers
$N^\text{SR,CR}_\text{prompt}(X)$ are predicted from simulated background.
The number $N^\text{CR}_\text{prompt}(\text{Rare})$ refers to simulated
background processes other than those being estimated, and includes single
top, DY, and diboson production. The number $N^\text{CR}_\text{nonprompt}$
refers to the estimate of the backgrounds with a nonprompt lepton from
data (as explained in Section~\ref{s:CCbkg}). Within the region defined by
the preselection, the CRs are obtained by requiring $\mathrm{BDT} < 0.2$ to get a
data sample enriched in background. They are further enriched in \ttbar
events by requiring them to have at least one tight \cPqb\ jet, and in
\wjets\ events by requiring the number of loose \cPqb\ jets to be zero. The
level of potential signal contamination in the CRs is well below 5\% and
is not expected to impact the final result.

The systematic uncertainties associated with these predictions are based
on differences between the predicted number of events (obtained from
Eq.~(\ref{eq:ddbckg})) and the observed number of data events, both in
validation regions as defined in the previous section.

The number $Y^\text{SR}_\text{nonprompt}$ of background events with
nonprompt leptons is estimated from data in all signal regions with the
method described in Section~\ref{s:CCbkg}. The yield of other SM processes
such as diboson, single top, and DY production are estimated from
simulation.

\subsection{Systematic uncertainties}
\label{s:MVAsys}

All processes that are modelled by simulation are subject to the same
systematic uncertainties as described in Section~\ref{s:CCsys}. The
statistical uncertainty of the signal simulation ranges between 3 and
11\%. The systematic uncertainty affecting the prediction of the
\wjets\ and \ttbar backgrounds has been described in
Section~\ref{s:MVAbckg}, where the statistical uncertainty from the number
of events in CRs is included. The uncertainties are evaluated from both
validation regions, and the larger value is conservatively chosen.
Furthermore, uncertainties on the shape of the BDT output, which can
affect the background prediction, have been assessed. They are smaller
than the aforementioned systematic uncertainties. The systematic
uncertainties affecting the prediction of the nonprompt lepton background
are the same as in Section~\ref{s:CCbkg}. As we perform a separate
analysis for each \dm\ region, the uncertainties are evaluated separately
and can therefore vary across different values of \dm. The relative
systematic uncertainties on the predictions of the \wjets, \ttbar, and
nonprompt lepton on the total background are provided in
Table~\ref{tab:mvasys}.

\begin{table}[!htbp]
\centering
\topcaption{The MVA search: relative systematic uncertainties (in \%) on the
          total background and signal prediction. The ``\NA'' means that a
          certain source of uncertainty is not applicable. In the case of the
          background, the uncertainties are on the total background. Systematic
          uncertainties on the data-driven prediction of the \wjets, \ttbar,
          and nonprompt lepton backgrounds are reported.}
\begin{tabular}{lcc}
\hline
Systematic  &     & \\
uncertainty & Background & Signal \\
\hline
Renormalization \&     &   &   \\
factorization scales    & 0--1  & 1.5--3.0  \\
Pileup                  & 1 &  1 \\
JES                     & 0--2  & 6--13   \\
JER                     & 0--1  & 1--6   \\
\cPqb-tagging           & 0--1  & 0--7    \\
Trigger                 & 1 &  1 \\
Lepton efficiency    & 0--1  & 4  \\
ISR (\ttbar and signal)                     & 0--1  & 0--6  \\
ISR (\wjets)             & 0--10  & \NA \\
\ptmiss modelling ({\sc FastSim})           & \NA & 5    \\
Prediction of \wjets                  & 7.1--32.0  & \NA \\
Prediction of \ttbar                  & 4.1--16.0  & \NA \\
Prediction of nonprompt               & 6.7--15.6  & \NA \\
\hline
\end{tabular}
\label{tab:mvasys}
\end{table}

\section{Results}
\label{s:res}

\begin{figure}
\centering
\includegraphics[width=1.00\textwidth]{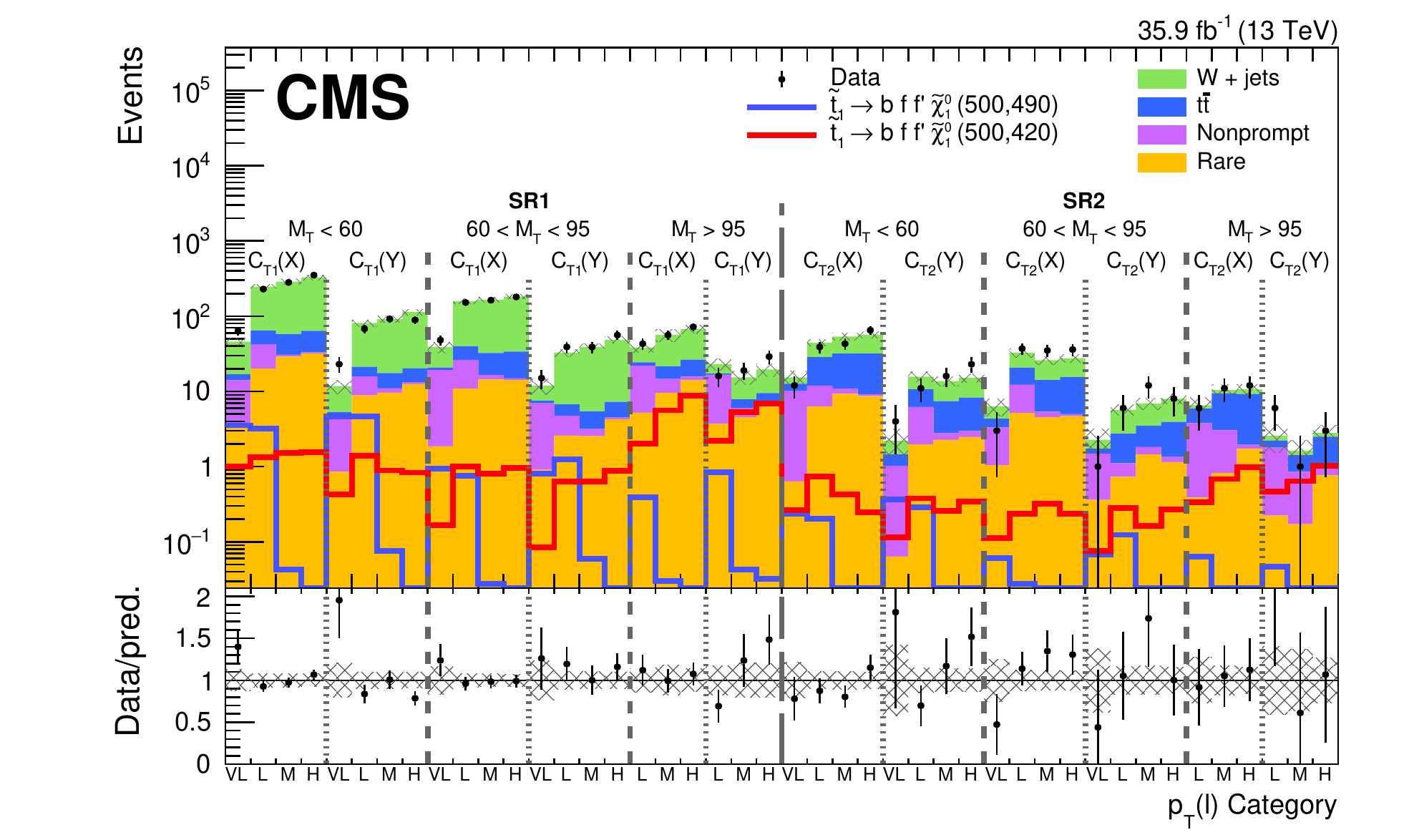}
\caption{The CC approach: summary of observed and expected background yields in all
        SRs as defined in Table~\ref{tab:CCsrcrdefs}. The vertical bars and the shaded
        areas represent the statistical uncertainty of the data and the total uncertainty
        in the prediction, respectively. The lower panel shows the ratio of data to
        prediction.
}
\label{fig:ccres}
\end{figure}

After performing the two searches, we find no evidence for direct top
squark production, as can be seen in Table~\ref{tab:mvares} and in
Fig.~\ref{fig:ccres} for the MVA and CC searches, respectively. Both sets
of results include the prediction of the \wjets\ and \ttbar processes, the
prediction of the background with a nonprompt lepton, the prediction of
other background processes from simulation, the total expected background,
and the observed number of data events. Systematic uncertainties are
included in the predictions. For the MVA search (Table~\ref{tab:mvares}),
the overlap between the SRs defined for different \dm\ is generally below
50\% for adjacent regions, and ranges from 0 to 30\% for nonadjacent
regions. Taking into account these results, the expected signal yield for
each \mmplane mass point, and the corresponding systematic uncertainties,
we interpret the absence of a clear excess in terms of a 95\% confidence
level (\CL) exclusion of top squark pair production in the \mmplane plane.
The limits are calculated according to the modified frequentist \CLs
criterion~\cite{frequentist_limit,Read:2002hq,LHC-HCG}. A test statistic,
defined to be the likelihood ratio between the background-only and
signal-plus-background hypotheses, is used to set exclusion limits on top
squark pair production. For the CC search, which features a larger number
of signal regions, an asymptotic approximation~\cite{cowan} is used, while
in the MVA search the distributions of these test statistics are
constructed using simulated experiments. Statistical uncertainties are
modelled as Poisson distributions. All systematic uncertainties are
modelled with a log-normal distribution. In the CC search, the effect of
signal contamination in the CRs is taken into account by including the
control regions, with the estimate of corresponding signal yields, in the
likelihood fit. When interpreting the results, we assume branching
fractions of 100\% for the two considered decay scenarios.

\begin{figure}[!htbp]
\centering
\includegraphics[width=0.67\textwidth]{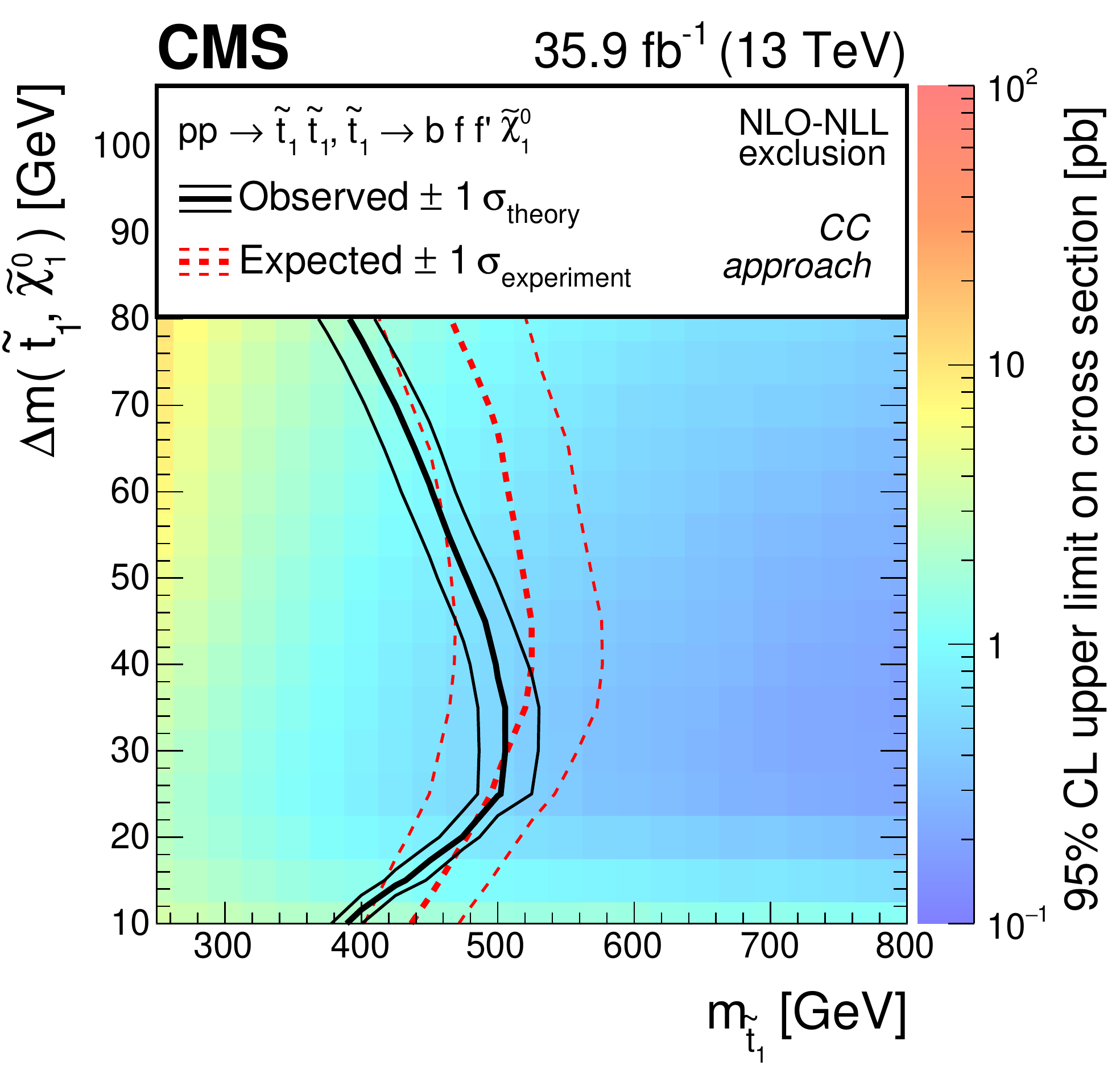} \\
\includegraphics[width=0.67\textwidth]{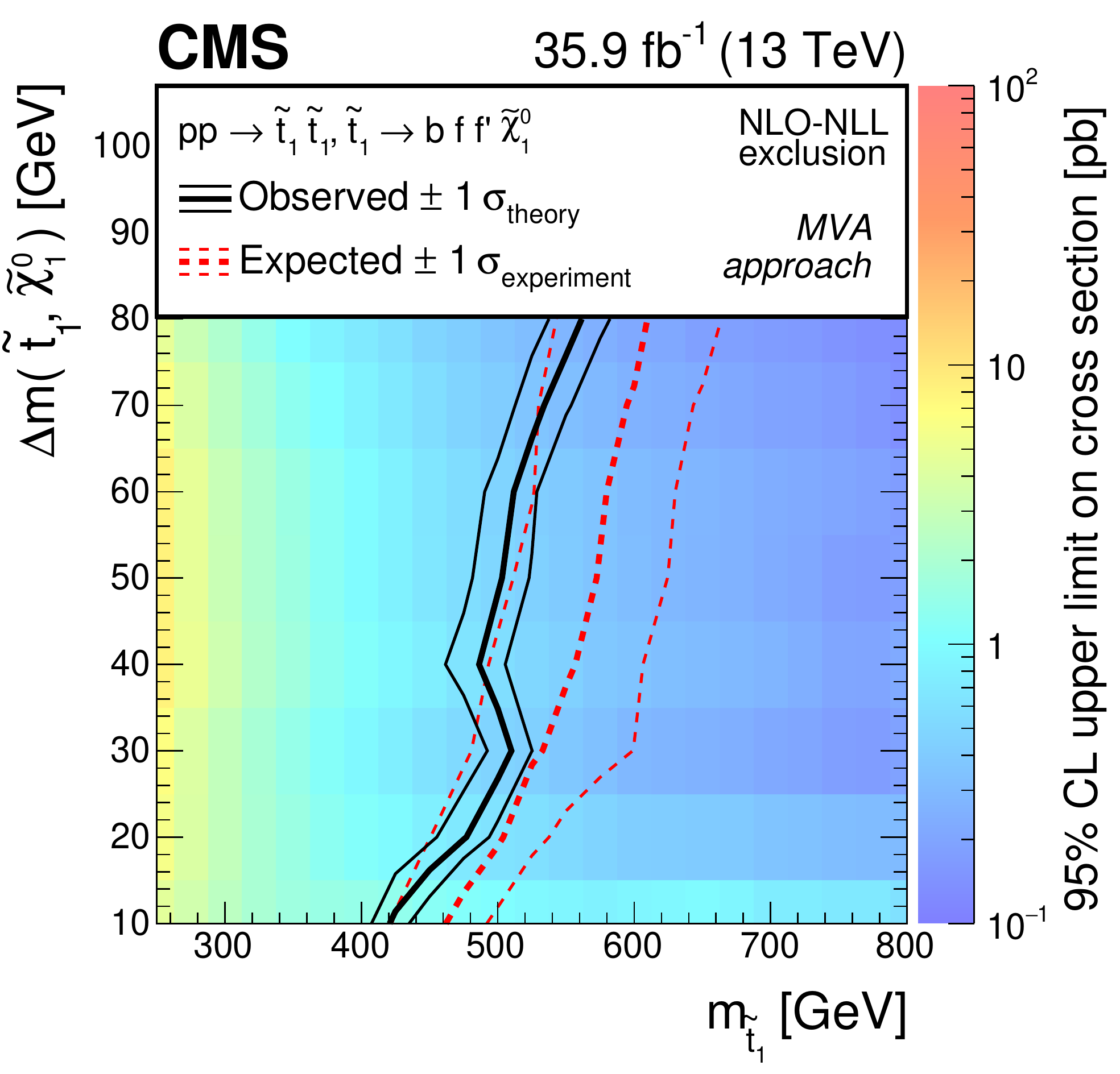}
\caption{Exclusion limit at 95\% \CL for the four-body decay of the top squark as a
        function of $m(\stp)$ and \dm\ for the CC (upper) and MVA (lower) approaches.
        The colour shading corresponds to the observed limit on the cross section.
        The solid black (dashed red) lines represent the observed (expected) limits,
        derived using the expected top squark pair production cross section. The
        thick lines represent the central values and the thin lines the variations due
        to the theoretical or experimental uncertainties.
}
\label{fig:1Lt2tt}
\end{figure}

Figure~\ref{fig:1Lt2tt} represents the exclusion contour as a function of
$m(\stp)$ and \dm\ for both searches in the case of the four-body decay
scenario.
For this decay mode, the CC search reaches its highest mass exclusion of 500\GeV for $\dm \approx 30\GeV$.
At $\dm = 80\GeV$, the analysis probes masses up to 390\GeV.
For the MVA search, the maximum sensitivity is reached for the highest
$\dm$ of 80\GeV, where top squark masses up to 560\GeV are excluded.
At $\dm = 10\GeV$, the corresponding value is 420\GeV.
For both analyses, the reduced sensitivity at the lowest mass differences is due to the decrease in the transverse momenta of the visible decay products, as shown in Fig.~\ref{fig:vardm}, and the corresponding loss in acceptance.
For intermediate values of $\dm$, the two approaches obtain similar limits.
At the highest mass differences, the MVA selection has higher acceptance than the CC approach as it also includes events with lepton $\pt>30\GeV$ while keeping the level of background under control.

\begin{figure}[!htbp]
\centering
\includegraphics[width=0.67\textwidth]{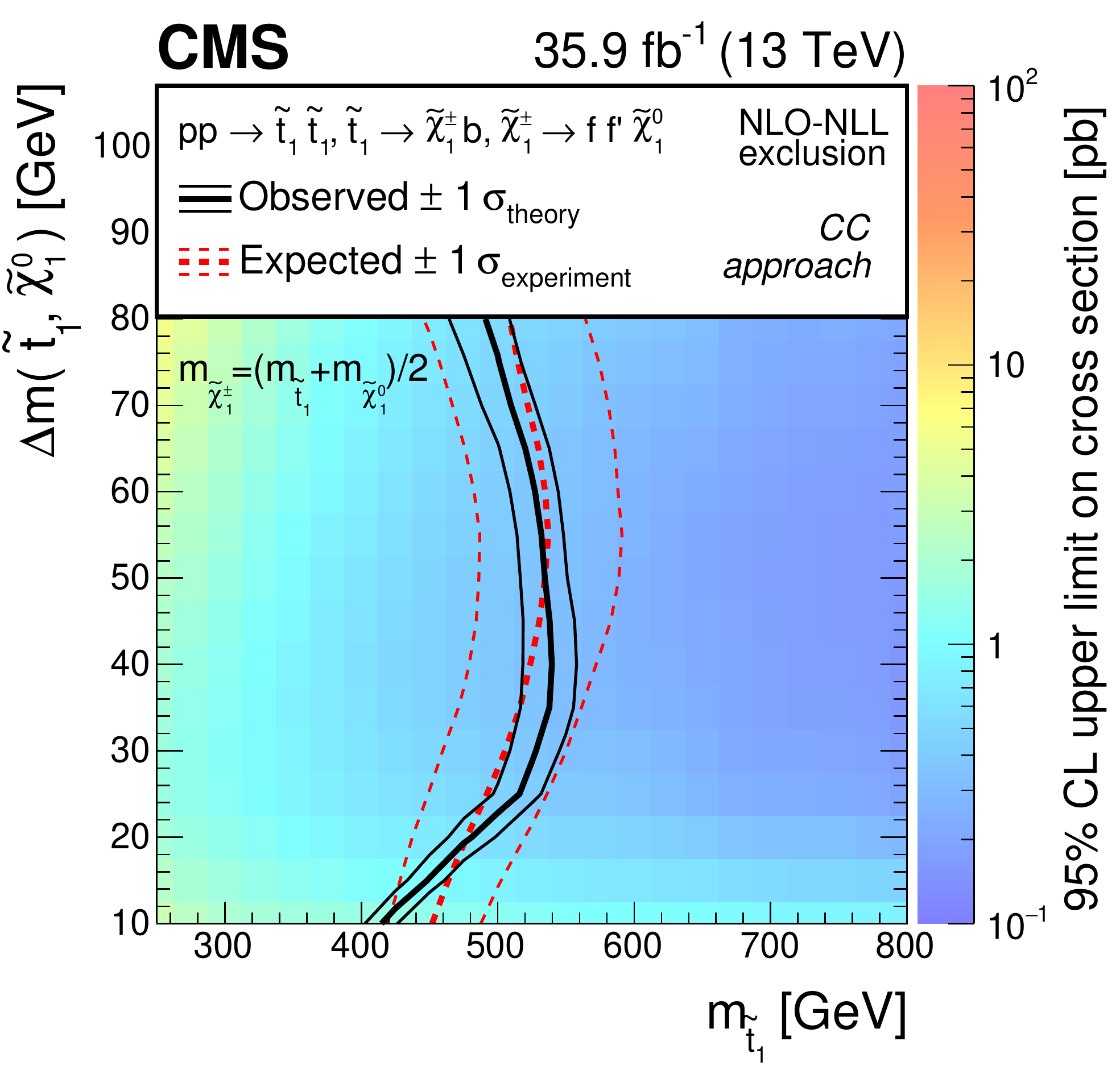}
\caption{Exclusion limit at 95\% \CL for the chargino-mediated decay of the top squark
        as a function of $m(\stp)$ and \dm\ for the CC search. The colour shading
        corresponds to the observed limit on the cross section. The solid black (dashed
        red) lines represent the observed (expected) limits, derived using the expected
        top squark pair production cross section. The thick lines represent the central
        values and the thin lines the variations due to the theoretical (experimental)
        uncertainties.
}
\label{fig:1Lt2bw}
\end{figure}

Figure~\ref{fig:1Lt2bw} represents the interpretation of the CC search for the chargino-mediated scenario.
For this model, with a chargino mass equal to the average of the top
squark and neutralino masses, the sensitivity is very similar to the
case of four-body decays, with the maximum exclusion being reached at slightly higher values of $\dm$.
Top squark masses of up to 540\GeV are excluded for $\dm \approx 40$\GeV.

\begin{figure}
\centering
\includegraphics[width=0.67\textwidth]{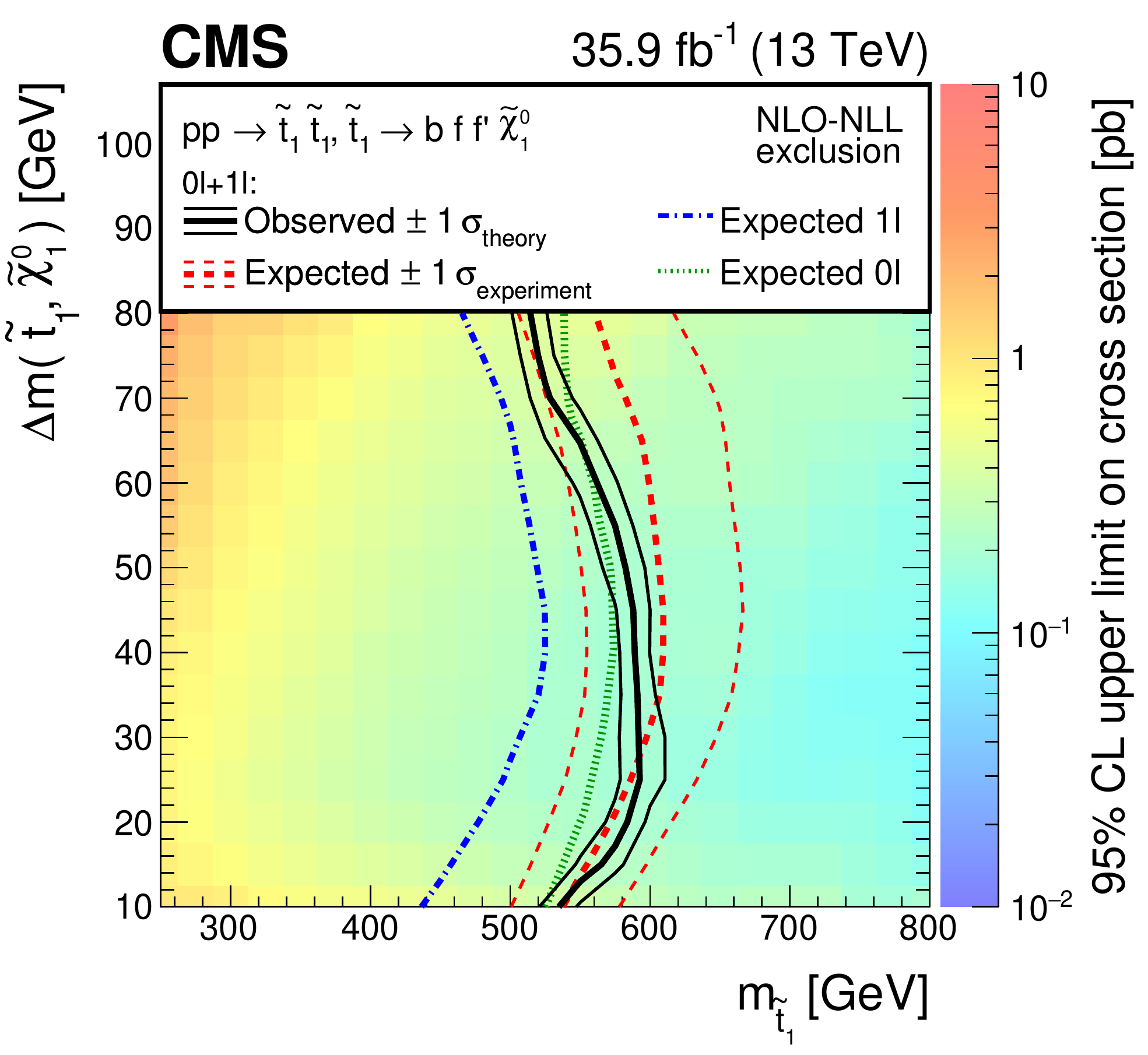}  \hfil
\includegraphics[width=0.67\textwidth]{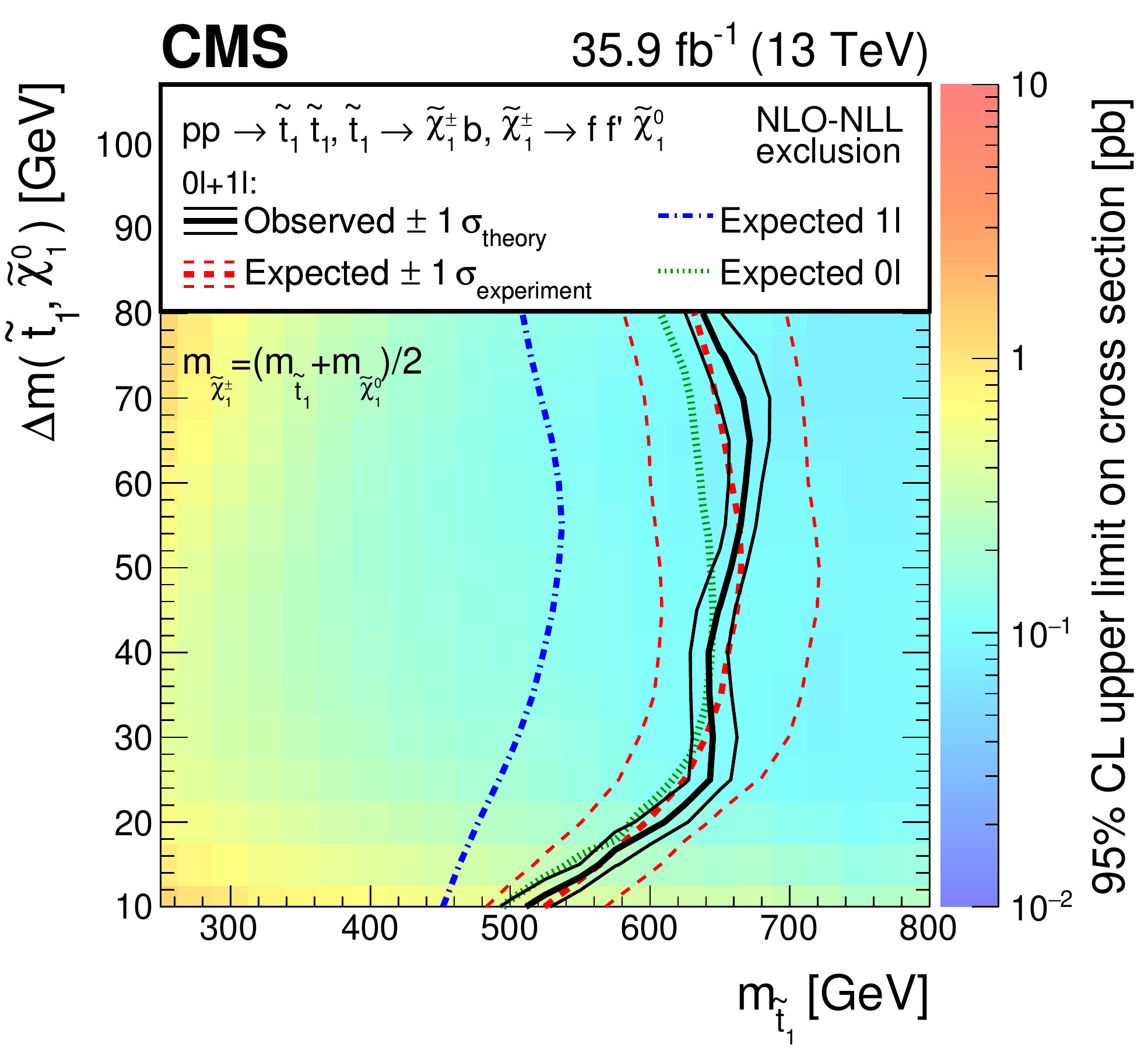} \\
\caption{Combined limits at 95\% \CL between the CC single-lepton ($1 \ell$) and all-hadronic
        ($0 \ell$)~\cite{Sirunyan:2017wif} searches for the four-body decay (upper) and the
        chargino-mediated decay (lower) of the top squark in the $m(\stp)$--$\Delta m(\stp,\PSGczDo)$
        plane. The correlations between the two searches have been taken into account.
        The colour shading corresponds to the observed limit on the cross section. The
        solid black (dashed red) lines show the observed (expected) mass limits, derived
        using the expected top squark pair production cross section. The thick lines
        represent the central values and the thin lines the variations due to the
        theoretical (experimental) uncertainties. The dot-dashed blue and dotted green
        lines show the individual expected mass limits for the $1 \ell$ and $0 \ell$
        searches, respectively.}
\label{fig:1L0Lcomb}
\end{figure}

In order to constrain top squark pair production in both decay modes using
the information from several final states, a statistical combination of
the CC search with the all-hadronic search~\cite{Sirunyan:2017wif} for
both decay scenarios of the top squark is performed. The common systematic
uncertainties of the two searches are treated as fully correlated, and the
possible correlations arising from events passing the selection criteria
of both searches are found to have negligible impact on the final results.
The combined limits, shown in Fig.~\ref{fig:1L0Lcomb}, include all SRs and
CRs of the all-hadronic and the single-lepton CC searches. The combination
of the two searches extends the exclusion limits on the top squark mass up to 590 and 670\GeV for the four-body and chargino-mediated scenarios, respectively.

\begin{table}[!htbp]
\centering
\topcaption{The MVA search: prediction of the \wjets, \ttbar, nonprompt lepton, and other
    backgrounds in the eight SRs defined by the threshold on the BDT output reported in
    the second column. The prediction of the first three processes is based on data, while
    that of $N^{SR}(\text{Rare})$, \ie rare backgrounds, is based on simulation. The
    uncertainties are the quadrature sum of the statistical uncertainties, the systematic
    uncertainties of Table~\ref{tab:mvasys}, and for the backgrounds predicted from
    simulation, the cross section uncertainties.
    The number of total expected background ($N^{SR}\text{(B)}$) and observed data
    ($N^{SR}\text{(D)}$) events in each SR are also reported.}
\begin{tabular}{lccccccc}
\hline
 & BDT$>$ & $Y^\text{SR}_\text{prompt}$ & $Y^\text{SR}_\text{prompt}$ &
$Y^\text{SR}_\text{nonprompt}$ & $N^\text{SR}$ & $N^\text{SR}\text{(B)}$ & $N^\text{SR}\text{(D)}$ \\
 &   & (\wjets) & (\ttbar) &  & (Rare) &  &  \\
\hline
$\dm$ = 10\GeV & 0.31 & 18.4 $\pm$ 3.6 & 1.8 $\pm$ 4.8 &  8.0 $\pm$ 2.9  & 2.3 $\pm$ 1.4 & 30.3 $\pm$ 6.7 & 39 \\
$\dm$ = 20\GeV & 0.39 & 9.0 $\pm$ 2.0 & 1.3 $\pm$ 1.7 & 11.2 $\pm$ 3.2  & 3.1 $\pm$ 1.9 & 24.7 $\pm$ 4.5 & 20 \\
$\dm$ = 30\GeV & 0.47 & 4.0 $\pm$ 2.5 & 1.2 $\pm$ 0.6 &  8.8 $\pm$ 2.5  & 1.7 $\pm$ 1.2 & 15.7 $\pm$ 3.7 & 22 \\
$\dm$ = 40\GeV & 0.48 & 4.1 $\pm$ 1.3 & 1.8 $\pm$ 0.7 &  7.6 $\pm$ 2.3  & 1.2 $\pm$ 0.9 & 14.8 $\pm$ 2.8 & 16 \\
$\dm$ = 50\GeV & 0.45 & 7.3 $\pm$ 2.1 & 4.7 $\pm$ 2.8 &  7.1 $\pm$ 2.0  & 5.5 $\pm$ 3.1 & 24.5 $\pm$ 4.8 & 36 \\
$\dm$ = 60\GeV & 0.50 & 2.0 $\pm$ 0.6 & 2.4 $\pm$ 1.2 &  3.1 $\pm$ 1.1  & 1.1 $\pm$ 0.9 &  8.7 $\pm$ 1.8 & 12 \\
$\dm$ = 70\GeV & 0.46 & 4.9 $\pm$ 1.6 & 3.4 $\pm$ 1.1 &  5.4 $\pm$ 1.6  & 3.2 $\pm$ 1.9 & 16.8 $\pm$ 2.9 & 20 \\
$\dm$ = 80\GeV & 0.44 & 7.1 $\pm$ 1.6 & 5.1 $\pm$ 0.9 &  5.3 $\pm$ 1.6  & 5.2 $\pm$ 3.0 & 22.8 $\pm$ 3.3 & 26 \\
\hline
\end{tabular}
\label{tab:mvares}
\end{table}

\section{Summary}
\label{s:conc}

A search for direct top squark pair production is performed in a
compressed scenario where the mass difference \dm\ between the lightest top
squark and the lightest supersymmetric particle (LSP), taken to be the
lightest neutralino \PSGczDo, does not exceed the \PW\ boson mass. Two
decay modes of the top squark are targeted: the four-body prompt decay to
$\cPqb \mathrm{f} \overline{\mathrm{f}}^{\,\prime} \PSGczDo$, and the
chargino-mediated decay to $\cPqb \PSGcpDo$ with a subsequent decay
$\PSGcpDo \to \mathrm{f} \overline{\mathrm{f}}^{\,\prime} \PSGczDo$. Results are
based on proton-proton collision data at $\sqrt{s}=13\TeV$, recorded
with the CMS detector in 2016 and corresponding to an integrated
luminosity of 35.9\fbinv. Selected events are required to have a
single lepton (electron or muon), and significant missing transverse
momentum (\ptmiss). Because of the small mass difference between the top
squark and the LSP, the decay products of the top squark are expected to
have low \pt. Events where the presence of a jet from initial-state
radiation leads to a boost of the top squark pair and sizeable \ptmiss are
selected.

Two search strategies are pursued. In the sequential selection approach
(CC), signal regions are defined based on discriminating variables,
particularly the transverse mass of the lepton-\ptmiss system and the lepton
momentum. In another approach, a multivariate analysis (MVA) is employed
that uses both kinematic and topological variables and is specifically
trained for different \dm\ regions of the four-body decay mode. In both
approaches, the dominant contributions to the signal regions from standard
model processes (\wjets, \ttbar, and events with misidentified leptons)
are estimated from control regions in data.

Data are found to be compatible with the predicted standard model
backgrounds. The results are used to set limits at 95\% confidence level
on the production cross section as a function of the \stp and \PSGczDo
masses, within the context of simplified models. Assuming 100\% branching
fraction in the decay channel under consideration and the top squark pair
production cross section computed at NLO+NLL
precision~\cite{prospino,Borschensky:2014cia,xs1,xs2,xs3,xs4,xs5}, these
limits are converted into mass limits.

Both search strategies are applied to the four-body decay mode. For this
decay mode, the MVA search excludes top squark masses up to 420 and
560\GeV at $\dm$ = 10 and 80\GeV, respectively.
The CC approach covers the chargino-mediated decays, where the chargino
mass is taken as the average of the top squark and the neutralino masses,
probing \stp masses up to 540\GeV for $\dm \approx 40$\GeV. The results of
the CC search have been combined with a search for top squark pair
production in the fully hadronic channel~\cite{Sirunyan:2017wif}. The
combined mass limits reach up to 590 and 670\GeV for four-body and
chargino-mediated decays, respectively. The reach of the \dm\ dependent
MVA search in the four-body decay mode is noteworthy, as the exclusion
limit goes beyond that of the combined result at high \dm.

The results summarized in this paper represent the most stringent limits
to date on the top squark pair production cross section for mass
differences between the top squark and the lightest neutralino below the
\PW\ boson mass, and for decays proceeding through the four-body or the
chargino-mediated modes.

\begin{acknowledgments}
We congratulate our colleagues in the CERN accelerator departments for the excellent performance of the LHC and thank the technical and administrative staffs at CERN and at other CMS institutes for their contributions to the success of the CMS effort. In addition, we gratefully acknowledge the computing centres and personnel of the Worldwide LHC Computing Grid for delivering so effectively the computing infrastructure essential to our analyses. Finally, we acknowledge the enduring support for the construction and operation of the LHC and the CMS detector provided by the following funding agencies: BMWFW and FWF (Austria); FNRS and FWO (Belgium); CNPq, CAPES, FAPERJ, and FAPESP (Brazil); MES (Bulgaria); CERN; CAS, MoST, and NSFC (China); COLCIENCIAS (Colombia); MSES and CSF (Croatia); RPF (Cyprus); SENESCYT (Ecuador); MoER, ERC IUT, and ERDF (Estonia); Academy of Finland, MEC, and HIP (Finland); CEA and CNRS/IN2P3 (France); BMBF, DFG, and HGF (Germany); GSRT (Greece); NKFIA (Hungary); DAE and DST (India); IPM (Iran); SFI (Ireland); INFN (Italy); MSIP and NRF (Republic of Korea); LAS (Lithuania); MOE and UM (Malaysia); BUAP, CINVESTAV, CONACYT, LNS, SEP, and UASLP-FAI (Mexico); MBIE (New Zealand); PAEC (Pakistan); MSHE and NSC (Poland); FCT (Portugal); JINR (Dubna); MON, RosAtom, RAS and RFBR (Russia); MESTD (Serbia); SEIDI, CPAN, PCTI and FEDER (Spain); Swiss Funding Agencies (Switzerland); MST (Taipei); ThEPCenter, IPST, STAR, and NSTDA (Thailand); TUBITAK and TAEK (Turkey); NASU and SFFR (Ukraine); STFC (United Kingdom); DOE and NSF (USA).

\hyphenation{Rachada-pisek} Individuals have received support from the Marie-Curie programme and the European Research Council and Horizon 2020 Grant, contract No. 675440 (European Union); the Leventis Foundation; the A. P. Sloan Foundation; the Alexander von Humboldt Foundation; the Belgian Federal Science Policy Office; the Fonds pour la Formation \`a la Recherche dans l'Industrie et dans l'Agriculture (FRIA-Belgium); the Agentschap voor Innovatie door Wetenschap en Technologie (IWT-Belgium); the F.R.S.-FNRS and FWO (Belgium) under the ``Excellence of Science - EOS" - be.h project n. 30820817; the Ministry of Education, Youth and Sports (MEYS) of the Czech Republic; the Lend\"ulet (``Momentum") Programme and the J\'anos Bolyai Research Scholarship of the Hungarian Academy of Sciences, the New National Excellence Program \'UNKP, the NKFIA research grants 123842, 123959, 124845, 124850 and 125105 (Hungary); the Council of Science and Industrial Research, India; the HOMING PLUS programme of the Foundation for Polish Science, cofinanced from European Union, Regional Development Fund, the Mobility Plus programme of the Ministry of Science and Higher Education, the National Science Center (Poland), contracts Harmonia 2014/14/M/ST2/00428, Opus 2014/13/B/ST2/02543, 2014/15/B/ST2/03998, and 2015/19/B/ST2/02861, Sonata-bis 2012/07/E/ST2/01406; the National Priorities Research Program by Qatar National Research Fund; the Programa Estatal de Fomento de la Investigaci{\'o}n Cient{\'i}fica y T{\'e}cnica de Excelencia Mar\'{\i}a de Maeztu, grant MDM-2015-0509 and the Programa Severo Ochoa del Principado de Asturias; the Thalis and Aristeia programmes cofinanced by EU-ESF and the Greek NSRF; the Rachadapisek Sompot Fund for Postdoctoral Fellowship, Chulalongkorn University and the Chulalongkorn Academic into Its 2nd Century Project Advancement Project (Thailand); the Welch Foundation, contract C-1845; and the Weston Havens Foundation (USA).
\end{acknowledgments}

\bibliography{auto_generated}
\cleardoublepage \appendix\section{The CMS Collaboration \label{app:collab}}\begin{sloppypar}\hyphenpenalty=5000\widowpenalty=500\clubpenalty=5000\vskip\cmsinstskip
\textbf{Yerevan Physics Institute, Yerevan, Armenia}\\*[0pt]
A.M.~Sirunyan, A.~Tumasyan
\vskip\cmsinstskip
\textbf{Institut f\"{u}r Hochenergiephysik, Wien, Austria}\\*[0pt]
W.~Adam, F.~Ambrogi, E.~Asilar, T.~Bergauer, J.~Brandstetter, E.~Brondolin, M.~Dragicevic, J.~Er\"{o}, A.~Escalante~Del~Valle, M.~Flechl, R.~Fr\"{u}hwirth\cmsAuthorMark{1}, V.M.~Ghete, J.~Hrubec, M.~Jeitler\cmsAuthorMark{1}, N.~Krammer, I.~Kr\"{a}tschmer, D.~Liko, T.~Madlener, I.~Mikulec, N.~Rad, H.~Rohringer, J.~Schieck\cmsAuthorMark{1}, R.~Sch\"{o}fbeck, M.~Spanring, D.~Spitzbart, A.~Taurok, W.~Waltenberger, J.~Wittmann, C.-E.~Wulz\cmsAuthorMark{1}, M.~Zarucki
% ...
\vskip\cmsinstskip
% ...
1:  Also at Vienna University of Technology, Vienna, Austria\\
% ...
\end{sloppypar}
\end{document}